\documentclass[twocolumn,superscriptaddress,amsmath,amssymb,showpacs,showkeys]{revtex4}

\usepackage{graphicx,color}
\usepackage{dcolumn}
\usepackage{bm}
\newcommand{\beq}{\begin{equation}} \newcommand{\eeq}{\end{equation}}
\newcommand{\bqa}{\begin{eqnarray}} \newcommand{\eqa}{\end{eqnarray}}

\definecolor{gold}{rgb}{0.75,0.56,0.00}
\definecolor{green}{rgb}{0.00,0.50,0.00}

\begin{document}

\preprint{APS/123-QED}

\title{Frequency Tracking and Parameter Estimation for Robust Quantum State-Estimation}

\author{Jason F. Ralph}
 \email{jfralph@liverpool.ac.uk} 
  \affiliation{ Department of Electrical Engineering and Electronics, The University of Liverpool,\\  Brownlow Hill, Liverpool, L69 3GJ, United Kingdom.}
 
 \author{Kurt Jacobs}
 \email{kurt.jacobs@umb.edu} 
 \affiliation{ Department of Physics, University of Massachusetts at Boston, 100 Morrissey Blvd, Boston, MA 02125, USA.}

\author{Charles D. Hill}
 \email{cdhill@unimelb.edu.au} 
 \affiliation{Centre for Quantum Computation and Communication Technology, School of Physics, University of Melbourne, Victoria 3010, Australia.}

\date{\today}

\begin{abstract} 
In this paper we consider the problem of tracking the state of a quantum system via a continuous weak measurement. If the system Hamiltonian is known precisely, this merely requires integrating the appropriate stochastic master equation. However, even a small error in the assumed Hamiltonian can render this approach useless. The natural answer to this problem is to include the parameters of the Hamiltonian as part of the estimation problem, and the full Bayesian solution to this task provides a state-estimate that is {\em robust} against uncertainties. However, this approach requires considerable computational overhead. Here we consider a single qubit in which the Hamiltonian contains a single unknown parameter. We show that classical {\em frequency estimation} techniques greatly reduce the computational overhead associated with Bayesian estimation and provide accurate estimates for the qubit frequency. 
\end{abstract}



\maketitle


\section{\label{sec:sec1}INTRODUCTION}

It is well known that the key to large scale quantum computing is the ability to maintain and manipulate a pure quantum state, minimizing any sources of possible error. Errors, either due to environmental fluctuations or inherent in the quantum system itself, are liable to produce decoherence, such as dissipation or dephasing, and lead to initially pure quantum states becoming mixed. In many of the candidate quantum technologies (such as quantum optics, atoms and ions in traps) the systems being used as the qubits can be reasonably well characterized. Laser frequencies can be measured accurately. Atomic and ionic transition frequencies are consistent from one qubit to another. In the solid state, where devices need to be fabricated individually, there will always be some uncertainty in the parameters that determine the Hamiltonians of the individual qubits, the coupling between qubits and the coupling to any external controls or environment. 

Any attempt to apply feedback control to solid-state qubits will therefore have to deal with uncertainties in their  Hamiltonians. Feedback control is based upon continuous state-estimation, in which the observer tracks the system via a continuous measurement~\cite{JacobsSteck06, Brun02, Bouten06,WisemanMilburnBook}, and uses this knowledge to correct the evolution via control inputs~\cite{DJ,James04,Ralph04,Jordan06}. It is therefore essential that continuous state-estimation handles uncertainties in the Hamiltonian. An estimation process that does this is referred to as being {\em robust}. An effective approach to achieving this robustness is to estimate continuously both the state of the system and the uncertain parameters in the Hamiltonian as the measurement proceeds. This is the approach we consider here -- comparing the optimal Bayesian technique and classical estimation techniques to determine an unknown parameter.

A number of authors have examined continuous state-estimation in which the Hamiltonian is uncertain. There has been work on parameter estimation by Gambetta and Wiseman~\cite{Gambetta01}, Verstraete \textit{et al.}~\cite{Verstraete01}, Stockton \textit{et al.}~\cite{Stockton04}, Chase and Geremia~\cite{Chase09}, and Tsang~\cite{Tsang09a,Tsang09b,Tsang10}.  Gambetta and Wiseman have examined a situation similar to the one we consider here, and derived the full Bayesian estimation equations. They found that the ability to determine the unknown parameter depended strongly on the way that the system was measured. This is a result of the way in which the measured observable is affected by errors in the parameter, and the noise introduced into the dynamics by the measurement. The goal of the work by Verstraete \textit{et al.} and Stockton \textit{et al.} was to estimate continuously a signal that appeared as a (possibly time dependent) parameter in the Hamiltonian of a measured system. This problem arises naturally in the context of measuring classical fields, such as gravity-wave detection~\cite{Braginsky95} and magnetometry~\cite{Stockton04}. In tracking the classical parameter it was necessary to track the state of the system, and thus the authors also derived the full Bayesian estimation equations. More recently, Chase and Geremia have developed a quantum analog of a classical tracking filter called a particle filter~\cite{Arulampalam02}, which estimates the evolution of continuous probability density function for a parameter by considering a finite set of weighted sample points~\cite{Chase09}, and Tsang has developed a generalisation of classical smoothing filters~\cite{Tsang09a,Tsang09b,Tsang10}, where measurements are processed forward and backwards in time to improve state estimates and the probability distributions for classical parameters. In addition, related work has been done by Yamamoto~\cite{Yamamoto06x}, who considered the problem of robust state-estimation for linear quantum systems, using results from classical control theory. Issues of robustness in quantum control have also been discussed in~\cite{Doherty00DC,DHelon06}.

We consider what is possibly the simplest non-linear quantum state-estimation problem, that of tracking the state of a single qubit. We first examine what happens when the observer uses the usual stochastic master equation to estimate the state of the system, but with an incorrect Hamiltonian. We find that this procedure is, as expected, very sensitive to errors in the Hamiltonian, failing quickly when the observer's Hamiltonian differs only a little from the true Hamiltonian. We then describe how this problem is solved by extending the estimation equations to include the Hamiltonian parameters. The resulting estimation equations are those derived in references~\cite{Gambetta01} and~\cite{Verstraete01}. (We also include a simple and concise derivation of these equations in an appendix.) Since the estimation equations simultaneously estimate both a quantum state and a set of classical parameters, we will refer to this set of equations as the {\em hybrid} stochastic master equation (HME). 

The HME, while an optimal solution to the robust state-estimation problem, does generate a considerable computational overhead. Our main purpose here is to show that this overhead can be reduced, with a relatively small reduction in performance, by using ``frequency tracking'' methods that were originally developed for classical signal processing applications (such as radar frequency tracking, sonar processing and communications)~\cite{Quinn01}. In particular, we examine three frequency tracking techniques: one Fourier-based method and two that do not require a Fourier transform (thereby avoiding a large computational overhead). The Fourier-based technique is the maximizer of the periodogram -- finding the frequency that maximizes the correlation between the noisy measurement signal and a complex Fourier kernel. It is optimal for a single sinusoidal signal in noise, but it is computationally expensive. 

The first non-Fourier technique (Quinn-Fernandes~\cite{Quinn91}) is iterative and can be applied entirely in the time domain. It uses notch filtering and relies on the iterative construction of a filter that cancels an instability at the desired frequency. It can be shown to approach the Cram\'{e}r-Rao bound for the convergence of the estimated frequency, in which the variance of the estimator asymptotes to $O(N^{-3})$, where $N$ is the number of data points in the measurement record~\cite{Quinn01}. 

The second non-Fourier technique is common in signal processing and is called `multiple signal characterisation' or MUSIC~\cite{Kaveh86,Kay88}. This technique uses the properties of the eigenvectors of the autocovariance matrix of the data. These are related to the corresponding eigenvalues, which are in turn related to the frequencies present in the data. The variance of this technique scales as $O(N^{-1})$ rather than $O(N^{-3})$, but -- unlike the Quinn-Fernandes method -- it does not require an initial frequency estimate and can be more stable for large, appropriately filtered data sets. We compare the accuracy of these techniques with that of the full Bayesian estimator and find that they perform very well in general, although some do require additional filtering to obtain a stable parameter estimate. 

We begin by reviewing the formalism used to describe continuous weak measurement,  and define the relevant measurement record upon which the parameter estimates will be based. We then describe the optimal Bayesian estimator (the HME) for updating the parameter estimates and discuss several approximate methods for the estimation of frequencies from noisy data, with particular emphasis on computational complexity. We then analyze the performance of the Fourier, Quinn-Fernandes and MUSIC techniques for a range of measurement strengths, and find that these perform well in comparison with the Bayesian method. We conclude by describing how these techniques can be included in a quantum feedback system to estimate an unknown (possibly time-dependent) Hamiltonian.


\section{\label{sec:sec2}CONTINUOUS WEAK MEASUREMENT AND STATE-ESTIMATION}

Measurement processes in quantum mechanics have always been the subject of intense theoretical interest. Traditionally, measurements have been projective and the quantum evolution has been reconstructed from a stochastic series of results, averaged over an ensemble of similarly prepared quantum systems. The main problem with this approach is the restrictive nature of the projective measurement process -- there is no alternative to performing a series of projective measurements and building up a probabilistic model for the average evolution. Continuous weak measurement models were developed as a means to describe the effect of weak (i.e. non-projective) measurement processes. This is achieved by coupling the quantum system of interest to environmental degrees of freedom, and then performing a series of projective measurements on the environment. Because the system becomes correlated with the environment, the projective measurements provide a stream of information about the system of interest. As long as the coupling strength between the system and the environment is sufficiently weak, then the effect of the projective measurements will be small compared to the underlying quantum evolution of the system, and the result is a (weak) continuous measurement. 

The evolution of the system now depends on the stream of measurement results, called the \textit{measurement record}, and since these are necessarily random, the evolution of the system is given by a stochastic equation. If one averages over all possible measurement records, then one obtains the usual Lindblad-form master equation for a system coupled to an environment. Each possible measurement record corresponds to a particular realisation of an experiment on a single quantum system. This is often referred to as an `unravelling' of the master equation, with different unravellings corresponding to different kinds of measurements that one could make on the environment~\cite{WisemanLinQ}. Weak measurements have a relatively small effect on the underlying quantum evolution, but increasing the strength of the coupling between the quantum system and the environment will also allow stronger measurements to be made -- including projective-like measurements, where the non-unitary terms dominate the evolution -- along with the quantum Zeno effect~\cite{Spiller94b}.

Of course, in a real experiment, all one has is the measurement record rather than the underlying processes. In such cases it is possible to reconstruct aspects of the underlying quantum evolution by solving the relevant stochastic master equation (SME) for the specific measurement process/unravelling using the measurement record obtained from the experiment. The resultant evolution provides estimates of the instantaneous state of the quantum system {\it given} the measurement record~\cite{DJ}. This type of state estimation process is common in classical systems, where the classical evolution of a system is estimated from a noisy series of measurements -- often using the Kalman filter or its variants~\cite{BarShalom01}. The parallels between classical state estimation and quantum state estimation have been widely discussed in the literature~\cite{DHJMT}. In this context, the estimated  quantum state represents the `best guess' of what the actual state of the system is, based upon the set of measurements that are currently available. As the system evolves and more information becomes available via the measurements, the estimate of the quantum state should improve and the purity of the state should increase.

For a quantum system with a Hamiltonian $\hat{H}$, subject to a measurement corresponding to an operator $\hat{c}$, the stochastic master equation
for the unravelling is given by~\cite{Wiseman05},
\begin{eqnarray}\label{sme1}
\hbar d\rho_{\mbox{\scriptsize c}}&=&-i\left[\hat{H},\rho_{\mbox{\scriptsize c}} \right]dt+\left(\hat{c}\rho_{\mbox{\scriptsize c}}\hat{c}^{\dagger}-\frac{1}{2}\left(\hat{c}\hat{c}^{\dagger}\rho_{\mbox{\scriptsize c}}+\rho_{\mbox{\scriptsize c}}\hat{c}\hat{c}^{\dagger}\right)\right)dt \nonumber\\
&&+ dz^{\dagger}\left(\hat{c}\rho_{\mbox{\scriptsize c}}-\left\langle\hat{c}\right\rangle\rho_{\mbox{\scriptsize c}}\right)+ \left(\rho_{\mbox{\scriptsize c}}\hat{c}^{\dagger}-\rho_{\mbox{\scriptsize c}}\left\langle\hat{c}^{\dagger}\right\rangle\right)dz
\end{eqnarray}
where $dt$ is an infinitesimal time increment and $dz$ is an infinitesimal complex Wiener increment. The complex Wiener increments obey the 
relations $E[dz] = 0$ (where $E[\ldots]$ represents an expectation value), $dz dz^{\dagger} = \hbar\Theta dt$, and $dz dz^T = \hbar\Upsilon dt$. In
general, $dz$ is a complex vector and $\Theta$ and $\Upsilon$ are matrices, but if we restrict consideration to a single measurement interaction, $\Theta$ is a real number (which we will set to one, corresponding to efficient detection~\cite{Wiseman05}) and we set $\Upsilon = 1$ so that the Wiener increments are real. The density matrix, $\rho_{\mbox{\scriptsize c}}$, is conditioned upon the measurement record, which is given by~\cite{Wiseman05}
\begin{equation}\label{meas1}
dy(t) = \left\langle \hat{c}^T\Theta+\hat{c}^{\dagger}\Upsilon \right\rangle_{\mbox{\scriptsize c}} dt + dz^T = \left\langle \hat{c}+\hat{c}^{\dagger} \right\rangle_{\mbox{\scriptsize c}} dt + dz
\end{equation}
where the expectation values are taken with respect to the conditional state, $\left\langle \hat{c}\right\rangle_{\mbox{\scriptsize c}} = Tr\left[\rho_{\mbox{\scriptsize c}} \hat{c}\right]$. 

Unlike projective measurements, weak measurement processes are not restricted to Hermitian operators but -- for the purposes of the current paper -- we will assume that the measurement operator is Hermitian. This means that the sole action of the measurement is to extract information about a single physical observable. (Measurement operators that are not Hermitian describe damping in addition to the extraction of information --- a common example is the damping of a lossy optical cavity~\cite{Spiller93}). For a Hermitian weak measurement $\hat{c} = \sqrt{2k\hbar}\hat{y} = \hat{c}^{\dagger}$, where $k$ is the strength of the measurement interaction, the SME reduces to,
\begin{eqnarray}\label{sme2}
d\rho_{\mbox{\scriptsize c}}&=&-\frac{i}{\hbar}\left[\hat{H},\rho_{\mbox{\scriptsize c}}\right]dt
-k\left[\hat{y}\left[\hat{y},\rho_{\mbox{\scriptsize c}} \right]  \right]dt \nonumber\\
&&+ \sqrt{2k}\left(\hat{y}\rho_{\mbox{\scriptsize c}}+\rho_{\mbox{\scriptsize c}}\hat{y}-2\left\langle\hat{y}\right\rangle_{\mbox{\scriptsize c}} \rho_{\mbox{\scriptsize c}} \right)dW
\end{eqnarray}
where $dW$ is a real Wiener increment (such that $dW^2 = dt$), and the measurement record is
\begin{equation}\label{meas2}
\frac{dy(t)}{\sqrt{\hbar}} = \sqrt{8k}\left\langle \hat{y} \right\rangle_{\mbox{\scriptsize c}} dt + dW
\end{equation}
At each point in time, the measurement record (from an experiment or a simulated process) will determine the associated $dW$, given the current estimate of the state $\rho_{\mbox{\scriptsize c}}$. The state of the system may then be updated using the value found for $dW$ and equation (\ref{sme2}). In classical state estimation, the value $dW$ is often referred to as the {\it innovation}~\cite{BarShalom01}, in that it represents the new information provided by the measurement. The innovation is normally the difference between the actual measurement (in this case, $dy(t)$) and the current estimate of what the measurement should be given the current state of knowledge of the system (which is $\sqrt{8k}\left\langle \hat{y} \right\rangle_{\mbox{\scriptsize c}} dt$). The main difference between classical and quantum measurement  processes is that if the measured observable does not commute with the Hamiltonian, then the measurement actually feeds noise into the observable that is being measured. Because of this, no matter how long we measure for, we will always be left with some uncertainty about the observable. 

The main potential problem with the above approach to state-estimation is that the stochastic master equation requires that the system is well-defined, in the sense that the Hamiltonian and the measurement interaction need to be known very accurately. If the Hamiltonian and interaction are not well defined, the measurement record that is used to update the system will produce incorrect state estimates and assign too much confidence to the results of the SME. This is also common in classical state estimation problems, where too much confidence is assigned to the dynamical processes and valid measurements are often discounted because they conflict with the (inaccurate) state estimates~\cite{BarShalom01}. Ideally, we would like our state estimates to be insensitive to small errors in the Hamiltonian, or to allow the parameters in the Hamiltonian to be estimated from the measurement record so that errors can be corrected on-line.

To examine the sensitivity of the state estimates to errors in the Hamiltonian, we consider two SMEs: one is used to produce a measurement record (playing the role of an experimental system) and the other is used to estimate the underlying state of the system based on a Hamiltonian that contains a small error. Initially, we select a simple single-qubit Hamiltonian,
\begin{equation}
H=\hbar\omega_x\sigma_x /2
\end{equation}
where $\sigma_x$ is the Pauli $x$ matrix. This Hamiltonian generates a rotation of the single qubit Bloch vector about the Bloch $x$-axis. We then apply an interaction that generates a measurement along the Bloch $z$-axis. This maximizes the size of the output signal relative to the noise by choosing a measurement that is orthogonal to the axis of rotation. The Hamiltonian and measurement are selected because they reflect a common situation for solid state charge qubits, where the $\sigma_x$ term corresponds to a tunneling interaction and the $\sigma_z$ interaction corresponds to a measurement of charge (see reference~\cite{Griffith06x} for example). Rewriting the SME in the Bloch vector representation~\cite{Scully97}, with
$$
r_i = Tr[\sigma_i \rho] \hspace{5mm} i = X,Y,Z , 
$$
where $\sigma_i$ is a Pauli matix and the density operator is
$$
\rho = \frac{I+r_X\sigma_y+r_Y\sigma_y+r_Z\sigma_y}{2} , 
$$
the SME becomes three coupled stochastic equations
\begin{eqnarray}\label{sme3}
dr_X &=&  - 4k r_X dt - \sqrt{8k} r_X r_Z dW , \nonumber \\
dr_Y &=&  -\omega_x r_Z dt - 4k r_Y dt - \sqrt{8k} r_Y r_Z dW , \\
dr_Z &=&  \omega_x r_Y dt + \sqrt{8k} (1-r_Z^2) dW . \nonumber
\end{eqnarray}
The corresponding measurement record is,
\begin{equation}\label{meas3}
\frac{dy(t)}{\sqrt{\hbar}} = \sqrt{8k}r_z dt + dW . 
\end{equation}
We use one SME with a fixed frequency $\omega_x$ to generate a measurement record, and then feed this measurement record to a second SME. This second SME is used by an observer to estimate the state of the system. The first SME is initialized in a pure state. The second SME is initialized in a completely mixed state ($r_X = r_Y = r_Z = 0$), reflecting the fact that the observer does not initially know what state the qubit is in. If the Hamiltonian used in the second SME is perfect (that is, is exactly the same as that used in the first SME), then the initial mixed state will gradually purify as information is extracted. It will converge to the corrrect underlying state. (The process of purification -- or state reduction -- of single qubits subject to weak measurements and the use of feedback to increase the rate of purification have been discussed in detail by the current authors and others \cite{rapidP,Combes06,Wiseman06x,Griffith06x}). If the Hamiltonian used in the second SME contains an error, the estimated/conditioned state will purify, but the errors in the predicted evolution due to the inaccurate Hamiltonian will cause the estimated state and the underlying state from the first SME to diverge. The question is how accurate the Hamiltonian needs to be so that the estimated state purifies before the prediction errors accumulate and cause the states to diverge. A robust parameter estimation process would allow a state to be purified to the required level before they diverge significantly. Preferably, this should be achieved without affecting the underlying evolution of the state, thereby allowing the Hamiltonian parameters to be estimated and updated. The alternative, a measurement strong enough to  completely dominate the evolution, projecting the state on a time-scale fast compared to the Hamiltonian, would certainly give the correct estimated state, but it is not the situation we are interested in here. 

\begin{figure}[htbp]\label{Fig_1}
	\centering
		\includegraphics[width=1.0\hsize]{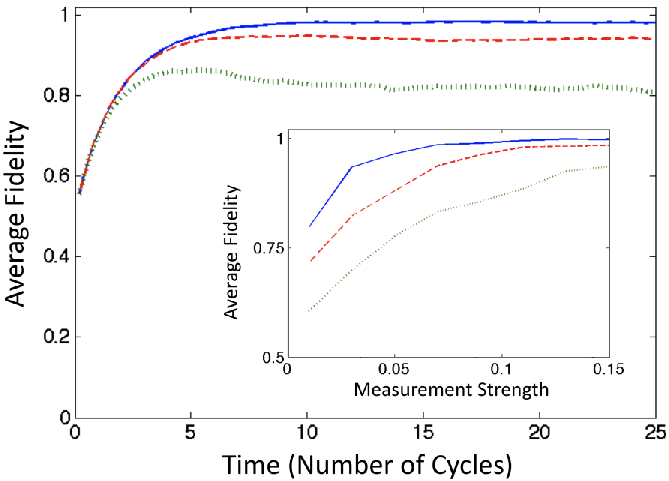}
	\caption{\label{fig:Average Fidelity} (Color online) Average fidelity of the estimated state vs. time for different values of expected errors $\sigma_{\delta\omega_x} = 1\%$ (blue-solid), $2\%$ (red-dash) and $5\%$ (green-dot) for $k = 0.07 \omega_x/(2\pi)$. Inset shows the saturated value for the average fidelity as a function of measurement strength $2\pi k/\omega_x$ for $\sigma_{\delta\omega_x} = 1\%$ (blue-solid), $2\%$ (red-dash) and $5\%$ (green-dot).}
\end{figure}

We begin by examining the divergence of the underlying state, given by the first SME, from that of the estimated state, given by the second SME. We do this by calculating the average fidelity between the two density matrices as a function of time for several different measurement strengths. The question relating to the extraction of the parameters in the Hamiltonian will be dealt with in the next section. The fidelity $F$ for two density matrices, $\rho_{0}$ and $\rho_{c}$, is a common measure of the difference between two (possibly mixed) density matrices and it is given by \cite{Peters04},
$$
F = F(\rho_{0},\rho_{c}) = \left|{\rm Tr}\left[\sqrt{\sqrt{\rho_{c}} \rho_{0} \sqrt{\rho_{c}}}\right]\right|^2
$$
In Fig.~\ref{Fig_1}, we show the average fidelity for a completely mixed initial state as a function of time for a number of different expected frequency errors (the oscillation frequency $\omega_x$ being the parameter that we will be trying to estimate and the errors being characterized by their standard deviation $\sigma_{\delta\omega_x}$ -- which will normally be given as a percentage) and for a measurement strength $k = 0.07 \omega_x/(2\pi)$. The fidelity is calculated by comparing the estimated density matrix $\rho_{\mbox{\scriptsize c}}$ and the underlying (pure state) density matrix $\rho_0$. Both SMEs are integrated numerically using Milstein stochastic integration~\cite{Kloeden92} and at least 4000 time-steps per oscillation cycle. The estimated states purify as information is extracted from the measurement record, thereby increasing their fidelity. The average purity of $\rho_{\mbox{\scriptsize c}}$ will increase with time (the estimated purity tends to one) but the average fidelity eventually saturates as the information gain from the measurement record is balanced by the information loss due to the inaccuracies in the estimated Hamiltonian parameter. In the inset figure in Fig.~1, the saturation value of the average fidelity is given as a function of measurement strength for three different values of the error ($\sigma_{\delta\omega_x} = 1\%, 2\%$ and $5\%$). Fig.~2 shows $1-\bar{F}$ ($1-$Average Fidelity, sometimes called the `infidelity') as a function of the percentage error in the estimate of the oscillation frequency for three different measurement strengths.
\begin{figure}[htbp]\label{Fig_2}
	\centering
		\includegraphics[width=1.0\hsize]{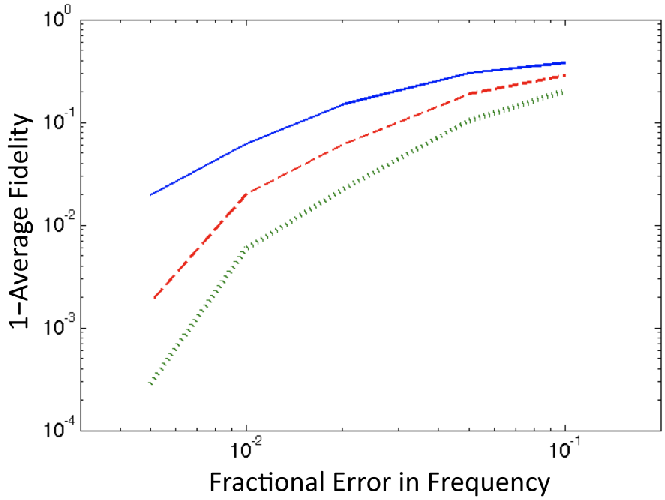}
	\caption{\label{fig:Average Fidelity Error} (Color online) Average fidelity error ($1-$Average Fidelity) of the estimated state vs. percentage frequency errors for $2\pi k/\omega_x = 0.03$ (blue-solid), $0.07$ (red-dash) and $0.12$ (green-dot) after 25 oscillation cycles.}
\end{figure}
It is clear from the data shown in Figure~1 and Figure~2 that errors of the order of a few percent can have a significant effect on the fidelity of the estimation process. To obtain a fidelity value in excess of $0.99$ the frequency error must be less than about $0.2-0.3\%$ if the measurement strength is around $k = 0.03\omega_x/2\pi$, but the error can be as large as $1\%$ if the measurement strength is greater than about $0.07-0.08\omega_x/2\pi$. Therefore, it is crucial that the measurement strength is chosen to reflect the expected accuracy of the frequency estimation process -- although, the stronger the coupling between the measurement environment and the qubit, the greater the effect of the back-action of the measurement on the qubit evolution.  In the next section, we consider how an optimal Bayesian inference method can be used to update the parameter estimates on-line and consider several alternatives taken from classical signal processing and frequency tracking.


\section{\label{sec:sec3}QUANTUM PARAMETER ESTIMATION AND FREQUENCY TRACKING}

\subsection{\label{sec:sec3pA} Bayesian Inference}

Making a continuous measurement will, in general, provide information not only about the state of a system, but also about the dynamics, and thus the parameters that determine the Hamiltonian. We will denote the parameters by the vector $\boldsymbol{\lambda}$. Our task is to process the measurement results in order to extract that information. The full solution to this problem is given by applying Bayesian inference~\cite{Jaynes03} (classical measurement theory) in combination with quantum measurement theory (being the quantum version of Bayesian inference~\cite{sk}). To begin with, our state of knowledge regarding the system and the parameters is described by a density matrix for the state of the system, $\rho_{\mbox{\scriptsize c}}$, and a probability density for the parameters, $P(\boldsymbol{\lambda})$. Here we are concerned only with a single parameter (the frequency of the oscillations), so we have a scalar parameter, $\lambda$. The solution to the continuous estimation problem tells us how to update both the density matrix and $P(\lambda)$ in each infinitesimal time-step, $dt$ --- that is, it gives differential equations for $\rho_c$ and $P(\lambda)$, and these are stochastic because they are driven by the measurement record. 

We now present the stochastic differential equations that solve the combined state and parameter estimation problem. While they look somewhat involved, they are quite simple to derive, and we give this derivation in Appendix~\ref{AppA}. Since the Hamiltonian is a function of $\lambda$, we will write it as $H(\lambda)$. The stochastic master equation for the density matrix is 
\begin{eqnarray}
  d \rho_{\mbox{\scriptsize c}} & = & \frac{-i}{\hbar} \left[\int P(\lambda) H(\lambda) d\lambda ,\rho_{\mbox{\scriptsize c}} \right] dt - k [\hat{y},[\hat{y},\rho_c]] dt  \nonumber \\ 
  &  & +  4k (\hat{y}\rho_{\mbox{\scriptsize c}} + \rho_{\mbox{\scriptsize c}} \hat{y}  - 2\langle \hat{y} \rangle_{\mbox{\scriptsize c}} \rho_{\mbox{\scriptsize c}}) (dr - \langle \hat{y} \rangle_{\mbox{\scriptsize c}} dt) , 
  \label{eq::hmeA}
\end{eqnarray}
where $dr$ is the measurement record, $dr = dy(t)/\sqrt{8k\hbar}$, $\hat{y}$ is the measured observable, and $\langle \hat{y} \rangle_{\mbox{\scriptsize c}} = \mbox{Tr}[\hat{y}\rho_{\mbox{\scriptsize c}}]$.  The equation of motion for $P(\lambda)$ is 
\begin{eqnarray}
  d P(\lambda)  & = &  8 k ( \langle \hat{y} \rangle - \langle \hat{y} \rangle_{\lambda}) ( dr -  \langle \hat{y} \rangle dt ) P(\lambda) . 
  \label{eq::hmeB}
\end{eqnarray}
Here we have introduced a new quantity, $\langle \hat{y} \rangle_{\lambda}$. This is the expectation value that $\hat{y}$ would have, {\em if} we knew the value of $\lambda$. To be able to calculate $\langle \hat{y} \rangle_{\lambda}$ we must continually update our state-of-knowledge of the system, assuming that we know $\lambda$, for every value of $\lambda$. Denoting our state-of-knowledge of the system, given $\lambda$, by $\rho_\lambda$, the update equation for each value of $\lambda$ is 
\begin{eqnarray}
  d\rho_\lambda & = & \frac{-i}{\hbar} [H(\lambda) ,\rho_\lambda]] dt - k [\hat{y},[\hat{y},\rho_\lambda]] dt \nonumber \\ 
  & &  + 4k (\hat{y}\rho_\lambda + \rho_\lambda \hat{y}  - 2 \langle \hat{y} \rangle_{\lambda} \rho) (dr - \langle \hat{y} \rangle_{\lambda} dt) 
  \label{eq::hmeC}
\end{eqnarray}
where $\langle \hat{y} \rangle_{\lambda} = \mbox{Tr}[\hat{y}\rho_\lambda]$. We will refer to Eqs.(\ref{eq::hmeA})-(\ref{eq::hmeC}) together as the \textit{hybrid master equation}, since it involves both quantum and classical states-of-knowledge. Unlike, the smoothing filters developed by Tsang~\cite{Tsang09a,Tsang09b,Tsang10}, the hybrid master equation propagates forwards in time only.

It is because we must propagate the density matrices $\rho_\lambda$ for every value of our parameter $\lambda$ that the full state-estimator is so numerically intensive. If $\lambda$ is a continuous variable, as is the case here, then it is impossible to have a density matrix for all values of $\lambda$, and we therefore use a discrete grid of values. Note that the spacing of values on this grid must be a least as small as the accuracy with which we wish to track $\lambda$. In addition, the length of the grid (the range of values of $\lambda$) must be large enough to encompass our initial uncertainty. If we use a fixed grid, these two requirements determine how many values of $\lambda$ require, which can be large. One approach to reducing this number would be to use a moving grid, updating the points on the grid along with $P(\lambda)$. Even this procedure would require at least three copies of the density matrix (likely more), and certainly more if we had two or more parameters. In addition, it requires the rather involved task of interpolating a new density matrix each time a grid point is moved. Rather than exploring dynamical grids as a numerical method for implementing the full estimator, in the next section, we consider replacing the Bayesian updater for $P(\lambda)$ with a classical frequency estimation technique. 

\begin{figure}[htbp]
	\centering
		\includegraphics[width=1.0\hsize]{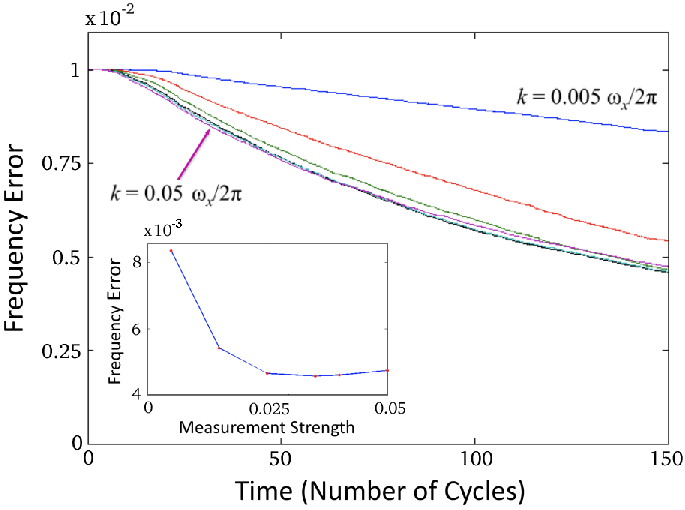}
	\caption{ (Color online) Expected Frequency Errors (1 standard deviation) vs. time for Bayesian Inference method and different measurement strengths, $2\pi k/\omega_x = 0.005, 0.015, 0.025, 0.035, 0.040 and 0.050$. Inset figure shows the expected frequency error after 150 qubit cycles for the data shown in the main figure.}
	\label{Fig_3}
\end{figure}

We present in Fig.~\ref{Fig_3} results for using the hybrid master equation to estimate the frequency of a single qubit. For these simulations we used a grid of $301$ points for $\lambda$. The simulations are highly intensive: running the estimator for $500$ periods, and averaging over $5000$ realizations, takes $\sim 33$ hours on a $232$ processor cluster. In Fig.~\ref{Fig_3} we plot the average error of the estimator (the average rms deviation from the true frequency) as a function of time, and for different values of the measurement strength $k$. While a stronger measurement provides a faster rate of information extraction, the laws of quantum mechanics mean that it also introduces more noise into the system. As a result there is an optimal value for the measurement strength for a given frequency. We see from Fig.~\ref{Fig_3} that this is approximately $k = 0.035 \omega / 2 \pi$, at least for a measurement time of $150$ cycles.  

\subsection{\label{sec:sec3pB} Classical Frequency Tracking}

The ability to determine which frequencies are present in a noisy signal is extremely important in a large number of practical engineering applications. In communication systems, analog signals are often transmitted by the modulation of the frequency of a carrier signal or digital data bits can be encoded in a frequency-shift key modulation scheme~\cite{FSKModulation}. In classical signal processing, the frequencies present in sonar signals provide information about the engine and gearing mechanisms in sea vessels~\cite{Quinn01} and tracking the Doppler frequency shifts in radar signals provides information about the relative (radial) motion of the radar and the target~\cite{DopplerRadar}. A large number of techniques have been developed to extract the dominant frequencies from a noisy signal. Some rely on calculating the Fourier transform of the data record and estimating the likelihood that frequency component corresponds to a signal rather than to noise, but these are computationally expensive and are similar in nature to the Bayesian method discussed above. However, other techniques have been developed which do not require explicit calculation of a Fourier transform and some can be applied in the time domain alone. It is natural then to ask whether these techniques can be used to extract useful information from a quantum measurement record of the form given in (\ref{meas3}). From the wide variety of techniques that are available, we have chosen three for further study: the maximizer of the periodogram, the Quinn-Fernandes method~\cite{Quinn91} and the {\em multiple signal characterisation} method or MUSIC~\cite{Kaveh86,Kay88}. The frequency that maximizes the periodogram provides the best correlation between the measurement record and a sine wave -- it is a computationally inefficient technique, but it provides the most accurate frequency estimate for cases where a single frequency signal is corrupted by white noise. The Quinn-Fernandes method is selected for its relative simplicity -- it works in the time domain, removing the need to perform computationally expensive Fourier transforms -- and for the fact that its theoretical accuracy approaches the optimal Cramer-Rao bound. The MUSIC method is slightly more complicated -- using an estimate of the signal's autocorrelation matrix -- but it does not require the initial frequency estimate used by Quinn-Fernandes and it can be used for signals containing more than one frequency component.

The main advantage of classical techniques is that the measurement record can be integrated over much longer time steps, thereby reducing the computational requirements. The optimal Bayesian inference technique requires that the SME is integrated using very small time steps. For the purity value of the quantum state to be well-behaved (i.e. not significantly larger than one), several thousand time steps are required per period of the oscillation even when Milstein stochastic integration techniques are used. By contrast, the classical tracking techniques only require on the order of fifty sample points per period -- a significant saving in computational terms. The measurement record used by the classical trackers is a discrete sequence of values sampled at regular intervals from the continuous measurement record. As such, each sample, $y_n$, is an integral of $dy(t)$ over a finite time-step $t_n\rightarrow t_n+\Delta t$ where $n = 0,... N-1$:
\begin{equation}
y_n = \int_{t_n}^{t_n+\Delta t}  dy(t)
\end{equation}
When the coupling between the qubit and the measured environment is very weak, the discrete signal $y_n$ can be approximated by a single sinusoid and a noisy background signal,
\begin{equation}\label{sine}
y_n \approx A\cos(\omega_x (n\Delta t)+\phi)+\epsilon_n
\end{equation}
where $A$ and $\phi$ are the amplitude and phase of the sinusoidal signal and $\epsilon_n$ is the noise. For a single sinusoid in Gaussian white noise, the minimum variance for an unbiased estimate of the frequency given by the asymptotic Cramer-Rao bound~\cite{Quinn01},
$$
\sigma^2_{CRB} = \frac{48\pi f_x(\omega_x)}{A^2(N\Delta t)^3}
$$
where $f_x(\omega)$ is the spectral density of the noise at a frequency $\omega$, and the signal-to-noise ratio at the qubit frequency is given by $SNR = 10\log_{10}(A^2/(4\pi f_x(\omega_x)))$~\cite{Quinn01}.

\subsubsection{Maximizing the periodogram}
The periodogram for $N$ discrete points is formed by calculating
\begin{equation}
I_y(\omega) = \left | \sum_{n = 0}^{N-1}y_n \exp(-i\omega t_n) \right|^2
\end{equation}
and the estimate of the qubit frequency is given by the value of omega that maximizes $I_y(\omega)$. For the cases considered in this paper, the number of points will be large and the signal-to-noise ratio for the signal/measurement record will be very low when compared to most signal processing applications. Typically, there will be fifty points per oscillation cycle of the qubit, and we will consider the performance over several hundred cycles. A rough estimate of the signal-to-noise ratio is given by the ratio between the variance of the two terms in the $dy(t)$ measurement record,
$$
SNR = 10 \log_{10}\left (\frac{(\sqrt{8 k} \Delta t)^2}{2 \Delta t}\right) = 10 \log_{10} (4 k \Delta t)
$$
which, for the discrete time examples that we will be considering, will provide signal-to-noise ratios between $-30$dB and $-20$dB. However, this simple calculation ignores the back-action of the measurement on the qubit, which induces slightly higher levels of noise in the output signal -- and broadens the peak associated with the sinusoidal oscillation of the qubit. In such cases, the accuracy of the Fourier-based technique is limited by the broadening of the peak because a broad peak has many maxima and a large number of points can generate a maximum when subject to a large amount of noise. Other techniques based on the Fourier transform -- such as the interpolation of Fourier coefficients~\cite{Huang00} -- offer better performance than simply maximizing the discrete Fourier transform of the discrete time signal (which was found to provide very poor performance for all the cases studied in this paper). However, these techniques will always under-perform when compared to maximizing the periodogram. An alternative approach based on fitting the whole frequency response of the broadened measurement peak could be envisaged, and could offer more accurate frequency estimates, but this is likely to require even more computational power than calculating the periodogram.

\subsubsection{The Quinn-Fernandes technique}

The Quinn-Fernandes technique belongs to a class called {\em adaptive notch filters}~\cite{Quinn91,Li98}. The general idea is to introduce a filter than removes a particular frequency from the signal, the `notch' frequency. If the signal is made up of white noise and a sinusoid, finding the notch frequency that minimizes the variance of the filtered signal is equivalent to finding the frequency of the sinusoid. The background to the method and the implementation of the technique are described fully in Appendix B. The main point to note is that this technique requires an initial estimate of the qubit frequency $\omega_1 = \tilde{\omega}_x$. The accuracy of the initial estimate will have an effect on the accuracy of the final frequency estimate $\omega_{QF}$ because it affects the convergence properties of the minimisation procedure used by the technique. For sinusoidal signals with low to moderate noise levels the Quinn-Fernandes technique is fairly insensitive to inaccuracies in this initial estimate, but for the situation considered here, where the signal-to-noise level is very low and the qubit oscillation is affected by the back-action of the measurement, the accuracy of the initial estimate is very important. Fig.~4 shows the evolution of the Quinn-Fernandes frequency estimates for the qubit signal for an example measurement record, and for several different initial frequency estimates. On short timescales there are very few stable solutions to the iterative procedure outlined above, but -- as time increases -- the accuracy of the initial estimate is required to be better because the number of stable solutions/estimates increases. This multipicity of solutions is partly due to the very low signal-to-noise level of the measurement record and partly due to the broadening of the corresponding peak in the Fourier domain (one effect of measuring the qubit is to reduce the effective quality factor of the oscillation). The Quinn-Fernandes technique performs much better when the noise level is (artifically) lowered. Reducing the size of the $dW$ term in the measurement record (equation~\ref{meas3}) relative to the size of the signal reduces the number of solutions, and the accuracy of the technique approaches the theoretical best-case. (The best-case corresponds to the output of the Quinn-Fernandes estimate when the initial estimate is perfect.) If the qubit signal is a spin-$\frac{1}{2}$ system, then moving to a larger value of angular momentum would boost the relative size of the signal compared to the $dW$ noise and may give the Quinn-Fernandes technique a significant advantage over the others. Increasing the signal-to-noise ratio by a factor of about 3 to 4 is enough to increase the accuracy of this technique significantly, and the Quinn-Fernandes technique is the least expensive computationally of the different techniques considered in this paper. Even without an increased signal-to-noise ratio, the Quinn-Fernandes does provide a slight improvement in accuracy over the initial estimate, as discussed in the next section. 
\begin{figure}[htbp]\label{Fig_4}
	\centering
		\includegraphics[width=1.0\hsize]{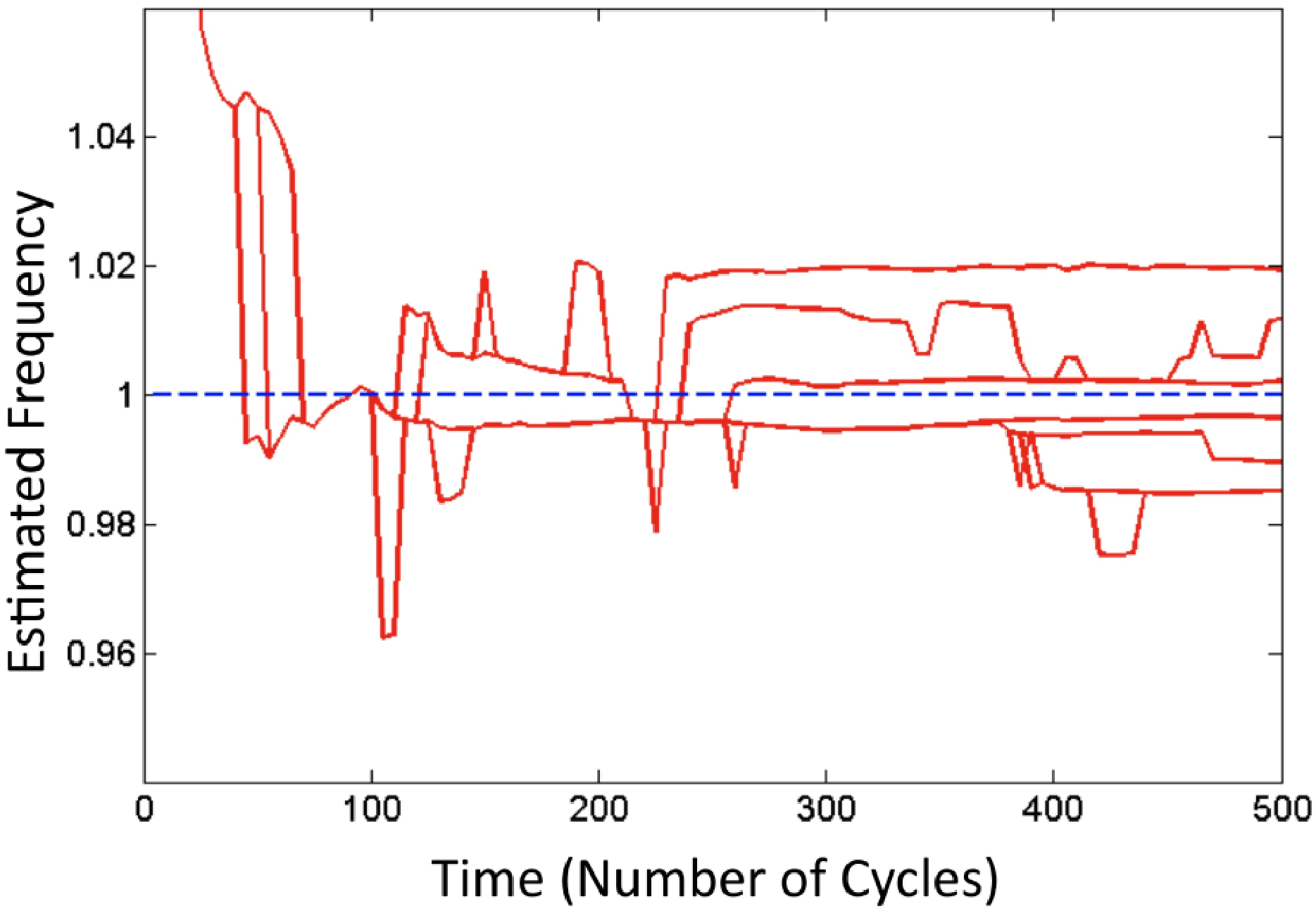}
	\caption{\label{fig:Q-F Frequency Example} (Color online)  Example outputs from Quinn-Fernades frequency estimation algorithm as a function of time with different initial frequency estimates: $ \tilde{f}_x = 1.02$ (red), $ \tilde{f}_x = 1.01$ (blue-x), $ \tilde{f }_x= 1.00$ (magenta-circle), $ \tilde{f}_x = 0.99$ (green-plus), and $ \tilde{f}_x = 0.98$ (purple-diamond).}
\end{figure}

\subsubsection{Multiple Signal Characterization (MUSIC)}
Multiple Signal Characterization, or MUSIC, is a widely used frequency estimation technique~\cite{Kaveh86,Kay88}. MUSIC uses properties of the eigenvectors of the autocovariance matrix of the measurement record to decompose the data into a signal sub-space and a noise sub-space. The autocovariance matrix is formed by constructing a symmetric Toeplitz matrix
\begin{equation}
C=\left(\begin{array}{cccc}
C_0 & C_1 & ... & C_{M-1} \\
C_1 & C_0 & ... & C_{M-2} \\
\vdots & \vdots & \ddots & \vdots\\
C_{M-1} & ... & C_1 & C_0 \\
\end{array}\right)
\end{equation}
from the autocovariances
\begin{equation}
C_j = N^{-1}\sum_{n=j}^{N-1}(y_n-\bar{y}) (y_{n-j}-\bar{y})
\end{equation}
where $\bar{y}$ is the mean of the measurement record. The size of the autocorrelation matrix $M > 2$ can be varied to optimize the performance of the technique and for the examples shown here $M=5$, which provides a good level of performance and does not require large computational resources. MUSIC can be used to find more than one frequency, however, we will only use it to find one frequency in this paper. 

Once the autocovariance matrix has been constructed, the eigenvectors of the matrix $\nu_m$ ($m=1...M$) are found and ordered so that the eigenvalues $\gamma_m$ are in ascending order $\gamma_1 \le \gamma_2 \le ... \le \gamma_M$. If there are $p$ sinusoids present in the data, the first $M-p$ eigenvectors corresponding to the lowest $M-p$ eigenvalues are used to find the MUSIC `spectrum' $S(\omega)$ from
\begin{equation}
S(\omega) = \frac{1}{\sum_{m=1}^{M-p}|e^*(\omega)\nu_{m}|^2}
\end{equation}
where $e(\omega) = [1\;  \exp(i\omega \Delta t) \;  \exp(2i\omega \Delta t) \; ... \; \exp(i(M-1)\omega \Delta t) ]^T$ and the eigenvectors $\nu_m$ are normalized. The MUSIC spectrum contains a number of maxima, one for each of the $p$ sinusoids present in the data. In our case, $p=1$ and $\omega \simeq \omega_x $. The maximum value of the MUSIC spectrum therefore provides our estimate for the qubit frequency. 

The signal-to-noise level in the measurement record is sufficiently high that the MUSIC technique applied on its own proves to be unstable -- the frequency estimates generated occasionally contain very large errors, which unduly affect the variance of the errors and the resultant accuracy of the technique. To remove these very large errors, it is necessary to pre-filter the measurement record to remove frequency components too far away from the initial estimate of the qubit frequency. This is not as restrictive as the initial estimate used in the Quinn-Fernandes technique: the pre-filter used here is a bandpass filter (a Butterworth filter~\cite{Butterworth}) with a pass band which is $\pm 10\%$ of the initial frequency estimate. Figure~5a shows an example of a MUSIC spectrum calculated for an example measurement record and the corresponding MUSIC frequency estimates are shown in Figure~5b as a function of time. 
\begin{figure}[htbp]\label{Fig_5}
	\centering
		\includegraphics[width=1.0\hsize]{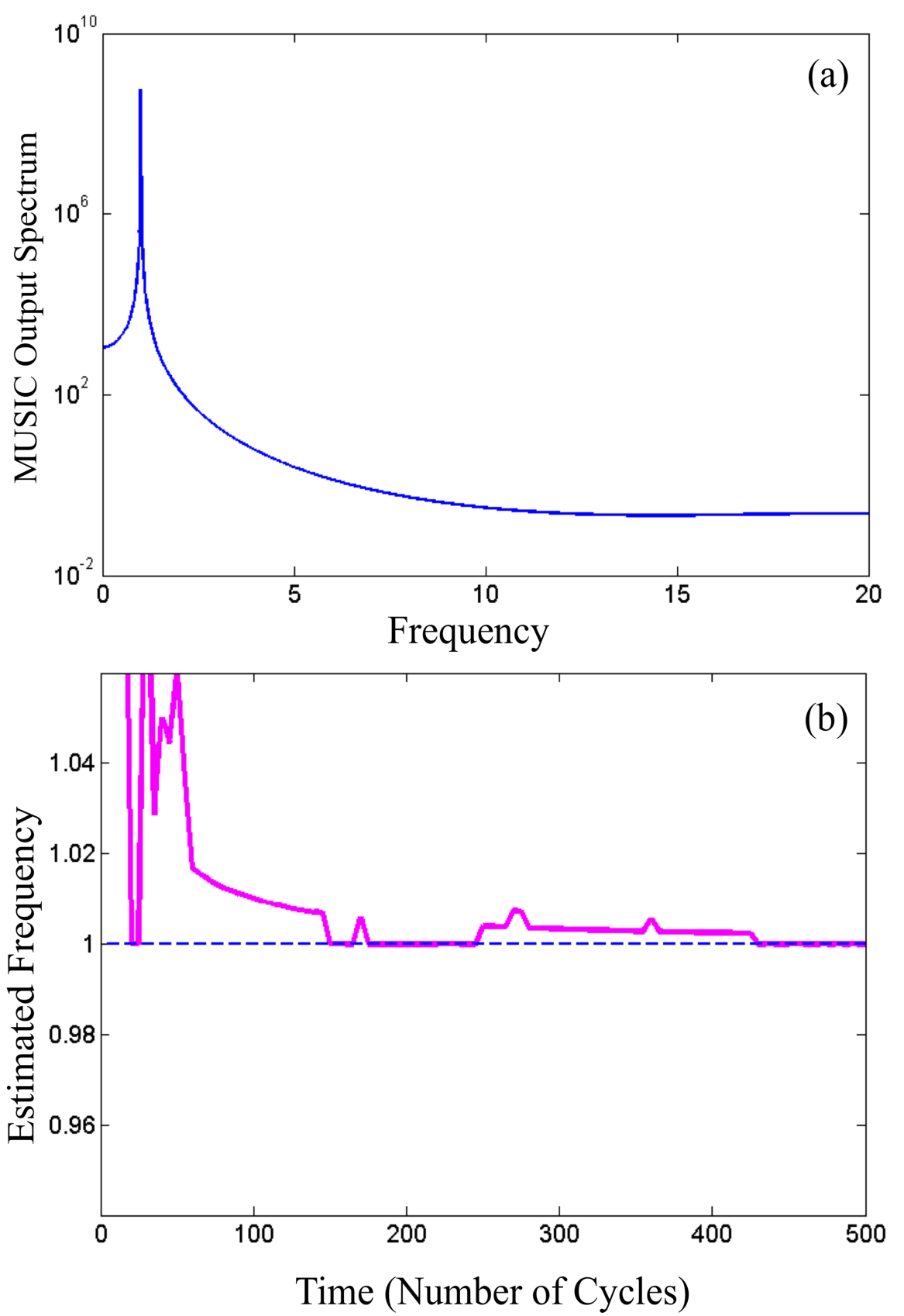}
	\caption{\label{fig:MUSIC Spectrum Example} (Color online) (a) Example MUSIC spectrum and (b) corresponding frequency estimate as a function of time.}
\end{figure}

\subsection{\label{sec:sec3pC} Comparison of Performance}

The principal aim of the frequency estimation algorithm is to improve the final fidelity of the estimated state in the stochastic master equation. Figures~1 and 2 show that the expected fidelity of the state, after the initial purification phase, is dependent on the measurement strength and the expected error in the qubit frequency. The Bayesian method gives the best estimates for the qubit frequency for relatively weak measurements, $2\pi k/\omega_x \simeq 0.035$. However, such weak measurements require very accurate frequency estimates to obtain high fidelity values: Figure~2 indicates that the frequency error should be less than about $0.2-0.3\%$ for the average fidelity to be in excess of $0.99$. Using stronger measurements, $2\pi k/\omega_x \simeq 0.070 - 0.120$, reduces the accuracy required from the frequency estimation algorithm to around $0.8-1\%$. The strength of the measurement interaction is an important design consideration for solid state circuits because it is normally more difficult to decouple or otherwise modify the interactions between fabricated circuits than it would be for an atomic or ionic qubit system. As such, the measurement strength would be an important design parameter in the construction of solid-state qubits and the associated control circuitry.

The accuracy of the frequency estimation algorithms for the different measurement strengths is summarized in Figure~6. For weak measurement values around $2\pi k/\omega_x \simeq 0.030$, the accuracy of each of the three classical techniques is found to be approximately equal in the long time limit, although maximizing the periodogram is more accurate over short timescales. (Two values are given for the Quinn-Fernandes technique, representing the upper and lower bounds to the accuracy -- which is a function of the accuracy of the initial estimate). The baseline accuracy of all three classical techniques is limited to around $0.8\%$, which would be accurate enough for a maximum average fidelity of around $F=0.95$ (see Figure~2). Only the Bayesian technique approaches the accuracy that would be required for fidelities approaching 0.99 -- at the expense of significantly increased computational loading.

As the measurement strength is increased, the accuracy of the periodogram/Fourier-based technique is seen to deteriorate relative to either of the other two classical techniques. For very short periods of time, it remains the most accurate technique, but the accuracy asymptotes very quickly to an error of around $1.2\%$. With an initial $1\%$ error in the estimate, the Quinn-Fernandes technique provides a slight improvement in the frequency estimate after about 150 qubit cycles. However, the accuracy of the Quinn-Fernandes technique for $2\pi k/\omega_x \simeq 0.070$ is around $0.9\%$, which is just outside the desired range for an average fidelity of $F=0.99$, and the accuracy of this approach can be seen to deteriorate when the measurement strength is increased. The lower bound for the Quinn-Fernandes accuracy is given in each graph in Figure~6 and is the accuracy of the Quinn-Fernandes technique if the initial estimate is perfect (i.e. zero error). This represents the best estimate that could be achieved using this (time-domain) technique, but in practice it will not be achievable because of errors in the initial frequency.

The expected accuracy of the MUSIC frequency estimates is slightly better than the other two classical methods, the error reduces to around $0.7\%$ at $2\pi k/\omega_x \simeq 0.070$. This is well within the range for an average fidelity greater than $0.99$. The expected accuracy of this technique would normally be expected to be worse than the other two methods because it does not converge to the Cramer-Rao bound in the long time limit, however, it is found to be remarkably robust in the presence of noise and it was also found to be stable in the presence of a broader peak in the Fourier domain. In fact, the expected accuracy was found to be fairly insensitive to noise and to variations in the measurement strength. It provides a reasonable estimate of the qubit frequency and is relatively efficient in terms of computational demands -- slightly longer to calculate than Quinn-Fernandes but much quicker that the periodogram or the Bayesian methods.

Whilst the Bayesian frequency estimates are most accurate in the region of $2\pi k/\omega_x \simeq 0.035$, the accuracy of this technique after 500 cycles is always good enough to give an average fidelity above 0.99. However, comparing the trends for the fidelity errors shown in Figure~2 with the accuracies of the frequencies given in Figure~6, it suggests that the improvement in the average fidelity for increasing measurement strength increases faster than the degradation in the frequency estimates. This means that the Bayesian technique  should still give higher fidelity estimates for the qubit states for stronger measurements even though frequency estimates are less accurate. It should also be noted that the Bayesian method generates a state estimate --- to obtain an estimate of the state from any of the classical tracking methods one must integrate the SME once the qubit frequency has been determined.
\begin{figure}[htbp]\label{Fig_6}
	\centering
		\includegraphics[width=1.0\hsize]{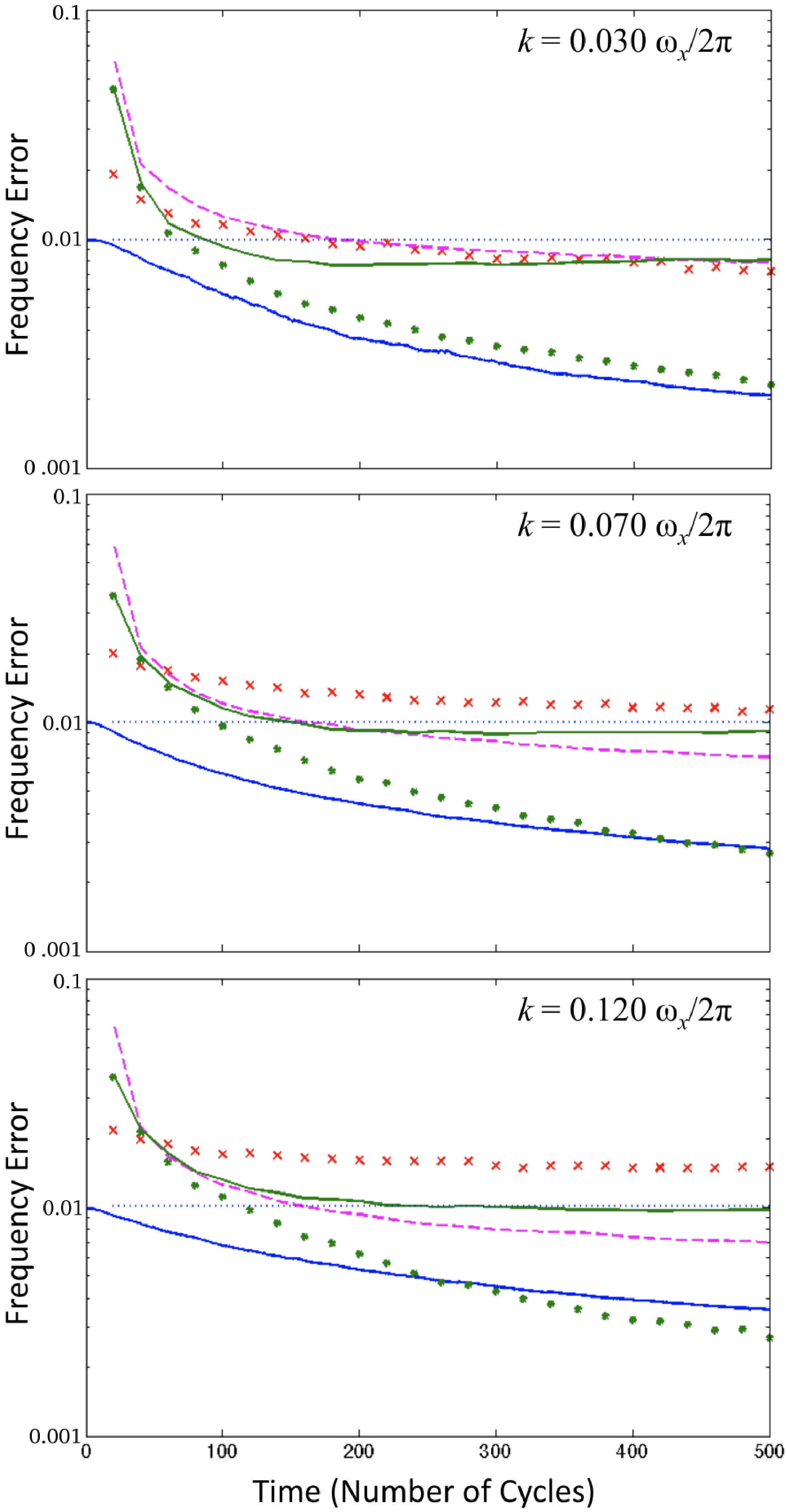}
	\caption{\label{fig:Frequency Tracking Errors} (Color online) Expected Frequency Errors (1 standard deviation) vs. time for classical frequency tracking techniques: Bayesian (blue-(lower) solid), the maximum of the periodogram (red-cross), Lower limit for Quinn-Fernades (green-dot) error, Upper Limit for Quinn-Fernandes (green-(upper) solid) with $1\%$ initial error (blue-dot), and MUSIC (purple-dash), for three different measurement strengths $2\pi k/\omega_x = 0.03$, $0.07$ and $0.12$}
\end{figure}

\section{\label{sec:sec4} FREQUENCY TRACKING FOR A TIME-DEPENDENT HAMILTONIAN} 

So far we have considered a frequency estimation process rather than a true tracking process, but the extension of the estimation processes discussed to tracking is relatively straightforward for the case of a single qubit subject to a continuous measurement. The simplicity derives from the fact that there is only one frequency to be estimated in the measurement signal. If the qubit frequency is fixed, then more measurement data can be collected and the accuracy of the frequency estimates can be improved beyond the values given in Figure~6. The accuracies of the Bayesian and MUSIC techniques are still improving after 500 qubit cycles. It is not clear whether this is also the case for the periodogram or Quinn-Fernandes techniques because any improvement in accuracy after a few hundred cycles is very slow in these cases. 

If the qubit frequency is not fixed but varies slowly with time -- slow compared with the timescales associated with the frequency estimation process -- there are a number of approaches that could be taken depending upon the nature of the variations in the frequency. For solid-state qubits, the most common problems are likely to be either a gradual frequency drift or a discontinuous jump. A gradual frequency drift could arise because the qubit controls are not absolutely stable and give rise to small drifts in the applied control fields leading to slight changes in the qubit frequency. Alternatively, a control could contain infrequent, but sudden, small jumps in voltage which alter the bias point of the qubit circuit in a random way. 

For gradual frequency drifts, a natural method for generating a time-dependent frequency `track' would be to generate a moving window for the measurement data so that the frequency is always estimated over a fixed length window of data where the frequency is approximately stationary. The estimates are continually updated as the window is moved. The length of the window can be varied depending upon the anticipated time-dependence of the underlying signal. This can be computationally expensive, the data having to be reprocessed each time the window moves, but it provides incremental updates to the frequency and generates a natural track history for the frequency. To reduce computational demands, the measurement data can be divided up into fixed blocks with finite length, but the frequency updates are then discrete measurements associated with each block and a separate tracking algorithm -- such as a Kalman filter~\cite{BarShalom01} -- is required to generate a continuous track in-between the estimates/updates. For sudden discrete frequency changes, there are a number of techniques based on hidden Markov models to generate the discontinuous transitions between different frequency values~\cite{Quinn01}. In each case, it is often helpful to have a model for the dynamics of the time-dependence of the system to reduce any residual errors in the tracked frequency values.

The frequency estimation techniques can be generalized if more than one frequency is present in the measurement signal. The MUSIC technique, in particular, is intended to allow multiple frequencies to be estimated. However, once measurements are made of more than one signal, potential problems can occur and slightly more thought is required when tracking multiple signals. When two frequencies are close together there are often problems deciding which measurement is associated with which tracked frequency. This is a well-known problem in object tracking. The mis-association of measurements with tracks can lead to significant errors in the parameter estimates. For algorithms like the Kalman filter, additional information regarding the expected errors in the parameters being tracked (the frequencies in this case) propagates with the estimates. The expected errors can then be used to minimize the probability of mis-association by finding the most likely association of measurements with tracks based on a statistically-weighted distance (e.g. the Mahalanobis distance~\cite{BarShalom01}). 


\section{\label{sec:sec5} CONCLUSIONS} 

We have considered a number of frequency estimation techniques that can be used with continuous quantum measurement processes. The techniques presented have been used to estimate the Hamiltonian for a single qubit, but they are quite general and can be readily adapted to monitor more complicated continuous measurement processes. The qubit has one parameter, the oscillation frequency for rotations around the Bloch $x-$axis. The most accurate estimates for the qubit frequency were generated by a Bayesian inference technique, but this technique is computationally expensive and typically requires thousands of integration steps per qubit cycle. By contrast, the three classical frequency estimation techniques considered -- maximized periodogram, Quinn-Fernandes, and MUSIC algorithms -- all performed relatively well using only 50 integration steps per qubit cycle and one (MUSIC) was found to be robust to measurement noise and produced accurate frequency estimates for a wide range of measurement strengths. The classical frequency estimation algorithms are much less computationally intensive that Bayesian inference, partly due to   the fact that they can be implemented with much larger time-steps. They provide frequency estimates that are accurate enough to generate robust (high fidelity) conditional state estimates for the qubit. We have also considered how these frequency estimates can be integrated into a tracking system for a time-dependent Hamiltonian, where the qubit frequency undergoes continuous drifts or sudden discontinuous changes.

\appendix

\section{Deriving the Hybrid Master Equation} 
\label{AppA}

The observer starts with a state of knowledge defined by her density matrix for the system, $\rho$, and her probability density for the parameters that define the Hamiltonian of the system. For simplicity we will assume that there is only one real parameter $\lambda$, so that the probability density is $P(\lambda)$. Generalizing the result to any number of parameters is straightforward. 

If the observer knew $\lambda$ precisely, then upon obtaining the measurement result $dr = \langle \hat{A} \rangle dt  + dW/\sqrt{8k}$, corresponding to a Hermitian operator $\hat{A}$, the evolution of the density matrix would be given by the stochastic master equation~\cite{JacobsSteck06}
\begin{eqnarray}
  d \rho_{\lambda} & = &  \frac{-i}{\hbar} \left[\hat{H} (\lambda)  ,\rho \right] dt - k [\hat{A} ,[\hat{A} ,\rho]] dt  \nonumber \\ 
     & & + 4k (\hat{A} \rho + \rho \hat{A}   - 2\langle \hat{A}  \rangle\rho) (dr - \langle \hat{A} \rangle dt) , 
  \label{eq::SMElam}
\end{eqnarray} 
so that the observers density matrix at time $t+dt$ is $\rho_{\lambda}(t+dt) = \rho + d\rho_{\lambda}$. 
Since  the observer does not know which value $\lambda$ takes, she must average over all the possible $\rho_{\lambda}(t+dt)$. Thus her state of knowledge at time $t+dt$ is actually  
\begin{eqnarray}
  \rho(t+dt) & = & \int_{-\infty}^{\infty} \!\!\!\!\!\! P(\lambda) \rho_\lambda(t+dt) d\lambda \nonumber \\ 
                  & = & \rho +  \int_{-\infty}^{\infty} \!\!\!\!\!\! \left[ P(\lambda) d\rho_\lambda \right]  d\lambda \nonumber \\ 
                  & \equiv &  \rho + d\rho .
  \label{eq::SMEapp}
\end{eqnarray}
The master equation for the observer's density matrix is therefore
\begin{eqnarray}
  d\rho & = &  \frac{-i}{\hbar} \left[ \int_{-\infty}^{\infty}  \!\!\!\!\!\!  P(\lambda) \hat{H} (\lambda) d\lambda ,\rho \right] dt - k [\hat{A} ,[\hat{A} ,\rho]] dt  \nonumber \\ 
     & & + 4k (\hat{A} \rho + \rho \hat{A}   - 2\langle \hat{A}  \rangle\rho) (dr - \langle \hat{A}  \rangle dt) 
\end{eqnarray} 
We now have the equation that updates the observer's density matrix, but we need also the equation that updates the observers probability density $P(\lambda)$. To derive this we must use Bayes' theorem~\cite{BT}. This tells us that the observer's new state of knowledge after obtaining the measurement result $dr$ is 
\begin{equation}
   P(\lambda | dr) = P(dr | \lambda) P(\lambda) / \mathcal{N} ,
\end{equation}
in which $P(dr | \lambda)$ is the probability density for $dr$ {\em given} $\lambda$, and $\mathcal{N}$ is a normalization constant.  If the value of the parameter was $\lambda$, then the measurement record would be given by $dr = \langle \hat{A}  \rangle_{\lambda} dt + dW/\sqrt{8k}$, where $\langle \hat{A}  \rangle_{\lambda} = \mbox{Tr}[\hat{A}  \rho_\lambda]$, and $\rho_\lambda$ is obtained by using the update equation Eq.(\ref{eq::SMElam}). Since $dW$ is Gaussian distributed with mean $0$ and variance $dt$, $P(dr | \lambda)$ is 
\begin{equation}
  P(dr | \lambda) = \frac{e^{-4k(dr - \langle \hat{A}  \rangle_{\lambda} dt)^2/dt}}{\sqrt{2\pi dt}} .
\end{equation}
The updated probability density is thus 
\begin{eqnarray}
  P(\lambda | dr) & \propto &  e^{-4k(dr - \langle \hat{A}  \rangle_{\lambda} dt)^2/dt} P(\lambda)   
\end{eqnarray}
where $dr$ is the actual measurement record obtained by the observer, $dr = \langle \hat{A}  \rangle dt  + dW/\sqrt{8k}$. We now expand this to first order in $dt$, remembering that $dW^2 = dt$~\cite{JacobsSteck06}. This gives 
\begin{eqnarray}
  P(\lambda | dr) & \propto & e^{-4k( \langle \hat{A}  \rangle dt - \langle \hat{A}  \rangle_{\lambda} dt + dW/\sqrt{8k})^2/dt} P(\lambda) \nonumber \\ 
     & \propto & e^{-4k( \langle \hat{A} \rangle_{\lambda} - \langle \hat{A}  \rangle)^2 dt + \sqrt{8k}( \langle \hat{A}  \rangle_{\lambda} - \langle \hat{A}  \rangle) dW} P(\lambda)  \nonumber \\ 
     & \propto & \left[ 1 + \sqrt{8k}( \langle \hat{A}  \rangle_{\lambda} - \langle \hat{A}  \rangle) dW \right]  P(\lambda)
\end{eqnarray}
The stochastic differential equation that updates the density $P(\lambda)$ is therefore 
\begin{eqnarray}
  d P(\lambda) & = & \sqrt{8k}( \langle \hat{A}  \rangle_{\lambda} - \langle \hat{A}  \rangle) dW P(\lambda) \nonumber \\ 
                       & = & 8k( \langle \hat{A}  \rangle_{\lambda} - \langle \hat{A}  \rangle) (dr - \langle \hat{A}  \rangle dt) P(\lambda) .
  \label{eq::dP}
\end{eqnarray} 
To be able to calculate this update to $P(\lambda)$ we must know $ \langle \hat{A}  \rangle_{\lambda}$ for {\em all} $\lambda$. To track the quantum state along with the parameter $\lambda$ we must therefore solve the master equation for $\rho$ (Eq.(\ref{eq::SMEapp})), the master equations for the $\rho_{\lambda}$ for every value of $\lambda$ (Eq.(\ref{eq::SMElam})), and the update equation for $P(\lambda)$, Eq.(\ref{eq::dP}), which is an example of a {\em Kushner-Stratonovich} equation.  

\section{Notch Filtering and the Quinn-Fernandes Method} 
\label{AppB}

For a discrete time series $x_n$, $n = 0,1,2,\ldots$, a filter $y_n$ can be constructed using a number of values for the input signal (feed-forward) and a number of previous values for the filter output (feedback). A general linear filter is given by:
\begin{eqnarray}\label{linearfilter}
y_n & = & a_0 x_n+a_1 x_{n-1}+ a_2 x_{n-2}+\ldots \nonumber\\
&&+ b_1 y_{n-1}+ b_2 y_{n-2}+\ldots
\end{eqnarray}
The feed-forward coefficients, $a_0, a_1,\ldots$, specify a finite duration impulse response -- i.e. if the input signal is a single impulse, the output signal returns to the default value (usually zero) after a finite number of time steps. Such a filter is normally referred to as a {\em finite impulse response} or FIR filter. The feedback coefficients, $b_1, b_2,\ldots$, define an {infinite impulse response} or IIR filter, because the output signal only returns to zero asymptotically. 

The properties of this type of filter are normally analyzed using the $z$-transform (more precisely, the {\em unilateral} $z$ transform)~\cite{Butterworth}, 
\begin{equation}
{\cal X}(z) = \sum_{n=0}^{\infty} x_n z^{-n}
\end{equation}
where $z$ is a complex number such that $z = |z|e^{i\omega}$, where $\omega$ is the (angular) frequency of the signal. Each factor of $z^{-1}$ represents a time delay of one time-step. The $z$-transform plays a similar role in the analysis of discrete time signals to that of the Laplace transform for continuous/analog signals. Applying the $z$-transform to (\ref{linearfilter}), we obtain
\begin{eqnarray}\label{linearfilter2}
{\cal Y}(z) & = & (a_0+a_1 z^{-1}+ a_2  z^{-1}+\ldots){\cal X}(z) \nonumber\\
&&+ (b_1 z^{-1}+ b_2  z^{-2}+\ldots){\cal Y}(z)
\end{eqnarray}
which can be rearranged to find the {\em transfer function} for the filter,
$$
{\cal H}(z) = \frac{{\cal Y}(z)}{{\cal X}(z)}=\frac{(a_0+a_1 z^{-1}+ a_2  z^{-1}+\ldots)}{(b_1 z^{-1}+ b_2  z^{-2}+\ldots)}=\frac{{\cal A}(z)}{{\cal B}(z)}
$$
The two polynomials ${\cal A}(z)$ and ${\cal B}(z)$, which represent the feed-forward and feedback parts of the filter respectively, contain all of the relevant properties for this type of linear filter. The zeros of ${\cal A}(z)$ determine which frequency components are suppressed by the filter -- if a zero of the polynomial lies on the unit circle in the $z$-plane, ${\cal A}(e^{i\omega_{0}})=0$, the frequency $\omega_0$ is removed. The zeros of ${\cal B}(z)$ give rise to poles in ${\cal H}(z)$ and determine the stability of the filter. For a stable filter, all of the poles of ${\cal H}(z)$ must lie within the unit circle, $|z|<1$. When a pole lies close to the unit circle, the frequencies close to the pole are enhanced.

The case considered in section~\ref{sec:sec3pB} consists of a single sinusoid and a noise source,
\begin{equation}\label{QF1a}
x_n = A\cos(n\omega \Delta t+\phi)+\epsilon_n
\end{equation}
for $n=0,... N-1$, where $A$ and $\phi$ are the amplitude and phase of the sinusoidal signal and $\epsilon_n$ is the noise. In the Fourier domain, this represents a single peak (from the sinusoid) and a large noisy background. The Quinn-Fernandes technique~\cite{Quinn91,Li98} introduces two filters -- one IIR filter and one FIR filter. The IIR filter has a pole which enhances one particular frequency, ${\cal B}(e^{i\omega})=0$. The FIR filter is designed to have zero response at the same frequency (the `notch' frequency, ${\cal A}(e^{i\omega})=0$). The variance of the output signal from the combined filter is a minimum when the filter frequency matches the frequency of the sinusoid. The Quinn-Fernandes algorithm provides a means to find the sinusoid frequency and, importantly for the computational efficiency, it operates purely in the time-domain. 

If the signal is given by (\ref{QF1a}), it should satisfy a filter equation,
\begin{equation}\label{QF2}
x_n - 2\cos(\omega \Delta t)x_{n-1}+x_{n-2} = \epsilon_n - 2\cos(\omega \Delta t)\epsilon_{n-1}+\epsilon_{n-2} 
\end{equation}
Rewriting Eq.(\ref{QF2}) as,
\begin{equation}\label{QF3}
x_n - \alpha x_{n-1}+x_{n-2} = \epsilon_n - \beta \epsilon_{n-1}+\epsilon_{n-2} 
\end{equation}
subject to the constraint $\alpha = \beta$, we start with an initial estimate $\omega_1$ for $\omega$, and set $\alpha_1 = 2\cos(\omega_1\Delta t)$ along with an index $j=1$. The data is then passed through the IIR filter to generate a new record $\zeta_{n,j}$
\begin{equation}\label{QF4}
\zeta_{n,j} = x_n - \alpha_j \zeta_{n-1,j}+\zeta_{n-2,j}
\end{equation}
where $\zeta_{n,j} = 0$ for $n < 0$. An improved estimate for $\beta$ is then constructed, by minimising the variance of the output from an FIR filter $e_n=\zeta_{n,j}-\beta\zeta_{n-1,j}+\zeta_{n-2,j}$ with respect to $\beta$, giving
\begin{equation}\label{QF5}
\beta_{j} = \frac{\sum_{n=0}^{N-1}(\zeta_{n,j}+\zeta_{n-2,j})\zeta_{n-1,j}}{\sum_{n=0}^{N-1}\zeta_{n-1,j}^2}
\end{equation}
If $|\alpha_j - \beta_j|$ is larger than some threshold then the $\beta$ value replaces the current $\alpha$ value, $\alpha_{j+1}=\beta_j$, and the process is applied again. If the difference between $\alpha$ and $\beta$ is small enough then the iteration in $j$ is stopped and the signal frequency is given as $\omega = \cos^{-1}(\frac{1}{2}\beta_j)/\Delta t$. In most situations the number of iterations in $j$ is fairly small, typically a few to ten iterations is sufficient to obtain a good frequency estimate, and the algorithm is also relatively insensitive to errors in the initial frequency estimate. However, these properties are normally quoted for systems with higher signal-to-noise ratios than those considered here, and that do not generate a back-action on the signal from the measurement process. In these slightly more benign situations the Quinn-Fernandes technique can be shown to generate errors which scale as $O(N^{-3})$ and approaches the Cramer-Rao bound, as long as the error in the initial estimate is fairly small, $O(N^{-1})$.

\begin{center}
	\textit{Acknowledgments}
\end{center}
\noindent  The authors would like to thank Prof.~B.~G.~Quinn for helpful and informative communications. This work was performed in part using the supercomputing facilities in the College of Science and Mathematics at UMass Boston. We are also grateful to Daniel Steck for allowing us to use his parallel cluster in the Oregon Center for Optics, University of Oregon. JFR and CDH would like to acknowledge the support of an ESPRC grant: EP/C012674/1. KJ is supported by the NSF under Project Nos. PHY-0902906 and PHY-1005571. This research is also partially supported by the ARO MURI grant W911NF-11-1-0268, and was performed in part using the supercomputing facilities in the School of Science and Mathematics at the University of Massachusetts Boston. CDH would also like to acknowledge the support of the Australian Research Council Centre of Excellence Scheme (CE110001027), the Australian Government, and the US National Security Agency (NSA) and the Army Research Office (ARO) under contract number W911NF-08-1-0527. 


\begin{thebibliography}{43}
\expandafter\ifx\csname natexlab\endcsname\relax\def\natexlab#1{#1}\fi
\expandafter\ifx\csname bibnamefont\endcsname\relax
  \def\bibnamefont#1{#1}\fi
\expandafter\ifx\csname bibfnamefont\endcsname\relax
  \def\bibfnamefont#1{#1}\fi
\expandafter\ifx\csname citenamefont\endcsname\relax
  \def\citenamefont#1{#1}\fi
\expandafter\ifx\csname url\endcsname\relax
  \def\url#1{\texttt{#1}}\fi
\expandafter\ifx\csname urlprefix\endcsname\relax\def\urlprefix{URL }\fi
\providecommand{\bibinfo}[2]{#2}
\providecommand{\eprint}[2][]{\url{#2}}

\bibitem[{\citenamefont{Jacobs and Steck}(2006)}]{JacobsSteck06}
\bibinfo{author}{\bibfnamefont{K.}~\bibnamefont{Jacobs}} \bibnamefont{and}
  \bibinfo{author}{\bibfnamefont{D.}~\bibnamefont{Steck}},
  \bibinfo{journal}{Contemporary Physics} \textbf{\bibinfo{volume}{47}},
  \bibinfo{pages}{279} (\bibinfo{year}{2006}).

\bibitem[{\citenamefont{Brun}(2002)}]{Brun02}
\bibinfo{author}{\bibfnamefont{T.~A.} \bibnamefont{Brun}},
  \bibinfo{journal}{{Am.\ J.\ Phys.}} \textbf{\bibinfo{volume}{70}},
  \bibinfo{pages}{719} (\bibinfo{year}{2002}).

\bibitem[{\citenamefont{Bouten et~al.}(2006)\citenamefont{Bouten, van Handel,
  and James}}]{Bouten06}
\bibinfo{author}{\bibfnamefont{L.}~\bibnamefont{Bouten}},
  \bibinfo{author}{\bibfnamefont{R.}~\bibnamefont{van Handel}},
  \bibnamefont{and} \bibinfo{author}{\bibfnamefont{M.}~\bibnamefont{James}},
  \bibinfo{journal}{{SIAM J. Control Optim.}} \textbf{\bibinfo{volume}{46}},
  \bibinfo{pages}{2199} (\bibinfo{year}{2007}).

\bibitem[{\citenamefont{Wiseman and Milburn}(2010)\citenamefont{Wiseman and Milburn}}]{WisemanMilburnBook}
\bibinfo{author}{\bibfnamefont{H.~M.} \bibnamefont{Wiseman}},
  \bibnamefont{and} \bibinfo{author}{\bibfnamefont{G.~J.}
  \bibnamefont{Milburn}}, \emph{\bibinfo{title}{Quantum Measurement and Control}}
  (\bibinfo{publisher}{CUP, Cambridge}, \bibinfo{year}{2010}).

\bibitem[{\citenamefont{Doherty and Jacobs}(1999)}]{DJ}
\bibinfo{author}{\bibfnamefont{A.~C.} \bibnamefont{Doherty}} \bibnamefont{and}
  \bibinfo{author}{\bibfnamefont{K.}~\bibnamefont{Jacobs}},
  \bibinfo{journal}{Phys.\ Rev.\ A} \textbf{\bibinfo{volume}{60}},
  \bibinfo{pages}{2700} (\bibinfo{year}{1999}).

\bibitem[{\citenamefont{James}(2004)}]{James04}
\bibinfo{author}{\bibfnamefont{M.~R.} \bibnamefont{James}},
  \bibinfo{journal}{Phys. Rev. A} \textbf{\bibinfo{volume}{69}},
  \bibinfo{pages}{032108} (\bibinfo{year}{2004}).

\bibitem[{\citenamefont{Ralph et~al.}(2004)\citenamefont{Ralph, Griffith,
  Clark, and Everitt}}]{Ralph04}
\bibinfo{author}{\bibfnamefont{J.~F.} \bibnamefont{Ralph}},
  \bibinfo{author}{\bibfnamefont{E.~J.} \bibnamefont{Griffith}},
  \bibinfo{author}{\bibfnamefont{T.~D.} \bibnamefont{Clark}}, \bibnamefont{and}
  \bibinfo{author}{\bibfnamefont{M.~J.} \bibnamefont{Everitt}},
  \bibinfo{journal}{Phys. Rev. B} \textbf{\bibinfo{volume}{70}},
  \bibinfo{pages}{214521} (\bibinfo{year}{2004}).

\bibitem[{\citenamefont{Jordan and Korotkov}(2006)}]{Jordan06}
\bibinfo{author}{\bibfnamefont{A.~N.} \bibnamefont{Jordan}} \bibnamefont{and}
  \bibinfo{author}{\bibfnamefont{A.~N.} \bibnamefont{Korotkov}},
  \bibinfo{journal}{Phys. Rev. B} \textbf{\bibinfo{volume}{74}},
  \bibinfo{pages}{085307} (\bibinfo{year}{2006}).

\bibitem[{\citenamefont{Gambetta and Wiseman}(2001)}]{Gambetta01}
\bibinfo{author}{\bibfnamefont{J.}~\bibnamefont{Gambetta}} \bibnamefont{and}
  \bibinfo{author}{\bibfnamefont{H.~M.} \bibnamefont{Wiseman}},
  \bibinfo{journal}{{Phys.\ Rev.\ A}} \textbf{\bibinfo{volume}{64}},
  \bibinfo{pages}{042105} (\bibinfo{year}{2001}).

\bibitem[{\citenamefont{Verstraete et~al.}(2001)\citenamefont{Verstraete,
  Doherty, and Mabuchi}}]{Verstraete01}
\bibinfo{author}{\bibfnamefont{F.}~\bibnamefont{Verstraete}},
  \bibinfo{author}{\bibfnamefont{A.~C.} \bibnamefont{Doherty}},
  \bibnamefont{and} \bibinfo{author}{\bibfnamefont{H.}~\bibnamefont{Mabuchi}},
  \bibinfo{journal}{Phys. Rev. A} \textbf{\bibinfo{volume}{64}},
  \bibinfo{pages}{032111} (\bibinfo{year}{2001}).

\bibitem[{\citenamefont{Stockton et~al.}(2004)\citenamefont{Stockton, Geremia,
  Doherty, and Mabuchi}}]{Stockton04}
\bibinfo{author}{\bibfnamefont{J.~K.} \bibnamefont{Stockton}},
  \bibinfo{author}{\bibfnamefont{J.~M.} \bibnamefont{Geremia}},
  \bibinfo{author}{\bibfnamefont{A.~C.} \bibnamefont{Doherty}},
  \bibnamefont{and} \bibinfo{author}{\bibfnamefont{H.}~\bibnamefont{Mabuchi}},
  \bibinfo{journal}{Phys. Rev. A} \textbf{\bibinfo{volume}{69}},
  \bibinfo{pages}{032109} (\bibinfo{year}{2004}).
  
\bibitem[{\citenamefont{Chase and Geremia}(2009)\citenamefont{Chase and Geremia}}]{Chase09}
\bibinfo{author}{\bibfnamefont{B.~A.} \bibnamefont{Chase}},
 \bibnamefont{and} \bibinfo{author}{\bibfnamefont{J.~M.} \bibnamefont{Geremia}},
  \bibinfo{journal}{Phys. Rev. A} \textbf{\bibinfo{volume}{79}},
  \bibinfo{pages}{022314} (\bibinfo{year}{2009}).

\bibitem[{\citenamefont{Tsang}(2009)\citenamefont{Tsang}}]{Tsang09a}
\bibinfo{author}{\bibfnamefont{M.}~\bibnamefont{Tsang}},
  \bibinfo{journal}{Phys. Rev. Lett.} \textbf{\bibinfo{volume}{102}},
  \bibinfo{pages}{250403} (\bibinfo{year}{2009}).

\bibitem[{\citenamefont{Tsang}(2009)\citenamefont{Tsang}}]{Tsang09b}
\bibinfo{author}{\bibfnamefont{M.}~\bibnamefont{Tsang}},
  \bibinfo{journal}{Phys. Rev. A} \textbf{\bibinfo{volume}{80}},
  \bibinfo{pages}{033840} (\bibinfo{year}{2009}).

\bibitem[{\citenamefont{Tsang}(2009)\citenamefont{Tsang}}]{Tsang10}
\bibinfo{author}{\bibfnamefont{M.}~\bibnamefont{Tsang}},
  \bibinfo{journal}{Phys. Rev. A} \textbf{\bibinfo{volume}{81}},
  \bibinfo{pages}{013824} (\bibinfo{year}{2010}).

\bibitem[{\citenamefont{Arulampalam et~al.}(2002)\citenamefont{Arulampalam, Maskell, Gordon, and Clapp}}]{Arulampalam02}
\bibinfo{author}{\bibfnamefont{M.~S.} \bibnamefont{Arulampalam,}},
  \bibinfo{author}{\bibfnamefont{S.} \bibnamefont{Maskell}},
  \bibinfo{author}{\bibfnamefont{N.} \bibnamefont{Gordon}},
  \bibnamefont{and} \bibinfo{author}{\bibfnamefont{T.}~\bibnamefont{Clapp}},
  \bibinfo{journal}{IEEE Trans. Sig. Proc.} \textbf{\bibinfo{volume}{50}},
  \bibinfo{pages}{174} (\bibinfo{year}{2002}).

\bibitem[{\citenamefont{Braginsky et~al.}(1995)\citenamefont{Braginsky,
  Khalili, and Thorne}}]{Braginsky95}
\bibinfo{author}{\bibfnamefont{V.~B.} \bibnamefont{Braginsky}},
  \bibinfo{author}{\bibfnamefont{F.~Y.} \bibnamefont{Khalili}},
  \bibnamefont{and} \bibinfo{author}{\bibfnamefont{K.~S.}
  \bibnamefont{Thorne}}, \emph{\bibinfo{title}{Quantum Measurement}}
  (\bibinfo{publisher}{CUP, Cambridge}, \bibinfo{year}{1995}).

\bibitem[{\citenamefont{Yamamoto}(2006)}]{Yamamoto06x}
\bibinfo{author}{\bibfnamefont{N.}~\bibnamefont{Yamamoto}},
  \bibinfo{journal}{Phys.\ Rev.\ A} \textbf{\bibinfo{volume}{74}},
  \bibinfo{pages}{032107} (\bibinfo{year}{2006}).

\bibitem{Doherty00DC}
\bibinfo{author}{\bibfnamefont{A.}~\bibnamefont{Doherty}},
  \bibinfo{author}{\bibfnamefont{J.}~\bibnamefont{Doyle}},
  \bibinfo{author}{\bibfnamefont{H.}~\bibnamefont{Mabuchi}},
  \bibinfo{author}{\bibfnamefont{K.}~\bibnamefont{Jacobs}}, \bibnamefont{and}
  \bibinfo{author}{\bibfnamefont{S.}~\bibnamefont{Habib}}, in
  \emph{\bibinfo{booktitle}{Proceedings of the 39th IEEE Conference on Decision
  and Control}}, vol.\ 1 
  (\bibinfo{publisher}{IEEE}, \bibinfo{address}{Washington, D.C.},
  \bibinfo{year}{2000}{\natexlab{a}}), p. \bibinfo{pages}{949}.

\bibitem[{\citenamefont{D'Helon and James}(2006)}]{DHelon06}
\bibinfo{author}{\bibfnamefont{C.}~\bibnamefont{D'Helon}} \bibnamefont{and}
  \bibinfo{author}{\bibfnamefont{M.~R.} \bibnamefont{James}},
  \bibinfo{journal}{Phys.\ Rev.\ A} \textbf{\bibinfo{volume}{73}},
  \bibinfo{pages}{053803} (\bibinfo{year}{2006}).

\bibitem[{\citenamefont{Quinn and Hannan}(2001)}]{Quinn01}
\bibinfo{author}{\bibfnamefont{B.}~\bibnamefont{Quinn}} \bibnamefont{and}
  \bibinfo{author}{\bibfnamefont{E.}~\bibnamefont{Hannan}},
  \emph{\bibinfo{title}{The Estimation and Tracking of Frequency}}
  (\bibinfo{publisher}{Cambridge, UK}, \bibinfo{year}{2001}).

\bibitem[{\citenamefont{Quinn and Fernandes}(1991)}]{Quinn91}
\bibinfo{author}{\bibfnamefont{B.~G.} \bibnamefont{Quinn}} \bibnamefont{and}
  \bibinfo{author}{\bibfnamefont{J.~M.} \bibnamefont{Fernandes}},
  \bibinfo{journal}{Biometrika} \textbf{\bibinfo{volume}{78}},
  \bibinfo{pages}{489} (\bibinfo{year}{1991}).

\bibitem[{\citenamefont{Kaveh and Barabell}(1986)}]{Kaveh86}
\bibinfo{author}{\bibfnamefont{B.~M.} \bibnamefont{Kaveh}} \bibnamefont{and}
  \bibinfo{author}{\bibfnamefont{A.~J.} \bibnamefont{Barabell}},
  \bibinfo{journal}{IEEE Trans. ASSP} \textbf{\bibinfo{volume}{34}},
  \bibinfo{pages}{331} (\bibinfo{year}{1986}).

\bibitem[{\citenamefont{Kay}(1988)}]{Kay88}
\bibinfo{author}{\bibfnamefont{S.~M.} \bibnamefont{Kay}},
  \emph{\bibinfo{title}{Modern spectral estimation: Theory and application}}
  (\bibinfo{publisher}{Prentince-Hall}, \bibinfo{year}{1988}).

\bibitem[{\citenamefont{Wiseman}(1996)}]{WisemanLinQ}
\bibinfo{author}{\bibfnamefont{H.~M.} \bibnamefont{Wiseman}},
  \bibinfo{journal}{Quantum Semiclass. Opt.} \textbf{\bibinfo{volume}{8}},
  \bibinfo{pages}{205} (\bibinfo{year}{1996}).

\bibitem[{\citenamefont{Spiller}(1994)}]{Spiller94b}
\bibinfo{author}{\bibfnamefont{T.~P.} \bibnamefont{Spiller}},
  \bibinfo{journal}{Phys. Lett. A} \textbf{\bibinfo{volume}{192}},
  \bibinfo{pages}{163} (\bibinfo{year}{1994}).

\bibitem[{\citenamefont{Bar-Shalom et~al.}(2001)\citenamefont{Bar-Shalom, Li,
  and Kirubarajan}}]{BarShalom01}
\bibinfo{author}{\bibfnamefont{Y.}~\bibnamefont{Bar-Shalom}},
  \bibinfo{author}{\bibfnamefont{X.-R.} \bibnamefont{Li}}, \bibnamefont{and}
  \bibinfo{author}{\bibfnamefont{T.}~\bibnamefont{Kirubarajan}},
  \emph{\bibinfo{title}{Estimation with Applications to Tracking and
  Navigation: Theory Algorithms and Software}}
  (\bibinfo{publisher}{Wiley-Interscience}, \bibinfo{year}{2001}).

\bibitem[{\citenamefont{Doherty
  et~al.}(2000{\natexlab{b}})\citenamefont{Doherty, Habib, Jacobs, Mabuchi, and
  Tan}}]{DHJMT}
\bibinfo{author}{\bibfnamefont{A.~C.} \bibnamefont{Doherty}},
  \bibinfo{author}{\bibfnamefont{S.}~\bibnamefont{Habib}},
  \bibinfo{author}{\bibfnamefont{K.}~\bibnamefont{Jacobs}},
  \bibinfo{author}{\bibfnamefont{H.}~\bibnamefont{Mabuchi}}, \bibnamefont{and}
  \bibinfo{author}{\bibfnamefont{S.~M.} \bibnamefont{Tan}},
  \bibinfo{journal}{Phys. Rev. A} \textbf{\bibinfo{volume}{62}},
  \bibinfo{pages}{012105} (\bibinfo{year}{2000}{\natexlab{b}}).

\bibitem[{\citenamefont{Wiseman and Doherty}(2005)}]{Wiseman05}
\bibinfo{author}{\bibfnamefont{H.~M.} \bibnamefont{Wiseman}} \bibnamefont{and}
  \bibinfo{author}{\bibfnamefont{A.~C.} \bibnamefont{Doherty}},
  \bibinfo{journal}{Phys.\ Rev.\ Lett.} \textbf{\bibinfo{volume}{94}},
  \bibinfo{pages}{070405} (\bibinfo{year}{2005}).

\bibitem[{\citenamefont{Spiller et~al.}(1993)\citenamefont{Spiller, Garraway,
  and Percival}}]{Spiller93}
\bibinfo{author}{\bibfnamefont{T.~P.} \bibnamefont{Spiller}},
  \bibinfo{author}{\bibfnamefont{B.~M.} \bibnamefont{Garraway}},
  \bibnamefont{and} \bibinfo{author}{\bibfnamefont{I.~C.}
  \bibnamefont{Percival}}, \bibinfo{journal}{Phys. Lett. A}
  \textbf{\bibinfo{volume}{179}}, \bibinfo{pages}{63} (\bibinfo{year}{1993}).

\bibitem[{\citenamefont{Griffith et~al.}(2007)\citenamefont{Griffith, Hill,
  Ralph, Wiseman, and Jacobs}}]{Griffith06x}
\bibinfo{author}{\bibfnamefont{E.~J.} \bibnamefont{Griffith}},
  \bibinfo{author}{\bibfnamefont{C.~D.} \bibnamefont{Hill}},
  \bibinfo{author}{\bibfnamefont{J.~F.} \bibnamefont{Ralph}},
  \bibinfo{author}{\bibfnamefont{H.~M.} \bibnamefont{Wiseman}},
  \bibnamefont{and} \bibinfo{author}{\bibfnamefont{K.}~\bibnamefont{Jacobs}},
  \bibinfo{journal}{Phys. Rev. B} \textbf{\bibinfo{volume}{75}},
  \bibinfo{pages}{014511} (\bibinfo{year}{2007}).

\bibitem[{\citenamefont{Scully and Zubairy}(1997)}]{Scully97}
\bibinfo{author}{\bibfnamefont{M.~O.} \bibnamefont{Scully}} \bibnamefont{and}
  \bibinfo{author}{\bibfnamefont{M.~S.} \bibnamefont{Zubairy}},
  \emph{\bibinfo{title}{Quantum Optics}} (\bibinfo{publisher}{Cambridge},
  \bibinfo{year}{1997}).

\bibitem[{\citenamefont{Jacobs}(2003)}]{rapidP}
\bibinfo{author}{\bibfnamefont{K.}~\bibnamefont{Jacobs}},
  \bibinfo{journal}{Phys. Rev. A} \textbf{\bibinfo{volume}{67}},
  \bibinfo{pages}{030301(R)} (\bibinfo{year}{2003}).

\bibitem[{\citenamefont{Combes and Jacobs}(2006)}]{Combes06}
\bibinfo{author}{\bibfnamefont{J.}~\bibnamefont{Combes}} \bibnamefont{and}
  \bibinfo{author}{\bibfnamefont{K.}~\bibnamefont{Jacobs}},
  \bibinfo{journal}{Phys. Rev. Lett.} \textbf{\bibinfo{volume}{96}},
  \bibinfo{pages}{010504} (\bibinfo{year}{2006}).

\bibitem[{\citenamefont{Wiseman and Ralph}(2006)}]{Wiseman06x}
\bibinfo{author}{\bibfnamefont{H.~M.} \bibnamefont{Wiseman}} \bibnamefont{and}
  \bibinfo{author}{\bibfnamefont{J.~F.} \bibnamefont{Ralph}},
  \bibinfo{journal}{New J. Phys.} \textbf{\bibinfo{volume}{8}},
  \bibinfo{pages}{90/1} (\bibinfo{year}{2006}).

\bibitem[{\citenamefont{Peters et~al.}(2004)\citenamefont{Peters, Wei, and
  Kwiat}}]{Peters04}
\bibinfo{author}{\bibfnamefont{N.~A.} \bibnamefont{Peters}},
  \bibinfo{author}{\bibfnamefont{T.-C.} \bibnamefont{Wei}}, \bibnamefont{and}
  \bibinfo{author}{\bibfnamefont{P.~G.} \bibnamefont{Kwiat}},
  \bibinfo{journal}{{Phys.\ Rev.\ A}} \textbf{\bibinfo{volume}{70}},
  \bibinfo{pages}{052309} (\bibinfo{year}{2004}).

\bibitem[{\citenamefont{Kloeden and Platen}(1992)}]{Kloeden92}
\bibinfo{author}{\bibfnamefont{P.~E.} \bibnamefont{Kloeden}} \bibnamefont{and}
  \bibinfo{author}{\bibfnamefont{E.}~\bibnamefont{Platen}},
  \emph{\bibinfo{title}{Numerical Solution of Stochastic Differential
  Equations}} (\bibinfo{publisher}{Springer}, \bibinfo{address}{Berlin},
  \bibinfo{year}{1992}).

\bibitem[{\citenamefont{Jaynes}(2003)}]{Jaynes03}
\bibinfo{author}{\bibfnamefont{E.~T.} \bibnamefont{Jaynes}},
  \emph{\bibinfo{title}{Probability Theory: The Logic of Science}}
  (\bibinfo{publisher}{CUP}, \bibinfo{address}{Cambridge},
  \bibinfo{year}{2003}).

\bibitem[{\citenamefont{Jacobs}(2002)}]{sk}
\bibinfo{author}{\bibfnamefont{K.}~\bibnamefont{Jacobs}},
  \bibinfo{journal}{Quantum Information Processing}
  \textbf{\bibinfo{volume}{1}}, \bibinfo{pages}{73} (\bibinfo{year}{2002}).

\bibitem[{\citenamefont{Lathi}(1998)}]{FSKModulation}
\bibinfo{author}{\bibfnamefont{B.~P.} \bibnamefont{Lathi}},
  \emph{\bibinfo{title}{Modern Digital and Analog Communication Systems}}
  (\bibinfo{publisher}{Oxford}, \bibinfo{year}{1998}).

\bibitem[{\citenamefont{Long et~al.}(1990)\citenamefont{Long, Mooney, and
  Skillman}}]{DopplerRadar}
\bibinfo{author}{\bibfnamefont{W.~H.} \bibnamefont{Long}},
  \bibinfo{author}{\bibfnamefont{D.~H.} \bibnamefont{Mooney}},
  \bibnamefont{and} \bibinfo{author}{\bibfnamefont{W.~A.}
  \bibnamefont{Skillman}}, in \emph{\bibinfo{booktitle}{Radar Handbook, 2nd
  Edition}}, edited by
  \bibinfo{editor}{\bibfnamefont{M.}~\bibnamefont{Skolnik}}
  (\bibinfo{publisher}{McGraw-Hill}, \bibinfo{year}{1990}).

\bibitem[{\citenamefont{Huang}(2000)}]{Huang00}
\bibinfo{author}{\bibfnamefont{D.}~\bibnamefont{Huang}},
  \bibinfo{journal}{Stat. Sinica} \textbf{\bibinfo{volume}{10}},
  \bibinfo{pages}{157} (\bibinfo{year}{2000}).

\bibitem[{\citenamefont{Li}(1998)}]{Li98}
\bibinfo{author}{\bibfnamefont{T.-H.}~\bibnamefont{Li}} \bibnamefont{and}
  \bibinfo{author}{\bibfnamefont{B.} \bibnamefont{Kedem}},
  \bibinfo{journal}{J. Time Series Analysis} \textbf{\bibinfo{volume}{19}},
  \bibinfo{pages}{69} (\bibinfo{year}{1998}).

\bibitem[{\citenamefont{Cavicchi}(2000)}]{Butterworth}
\bibinfo{author}{\bibfnamefont{T.~J.} \bibnamefont{Cavicchi}},
  \emph{\bibinfo{title}{Digital Signal Processing}}
  (\bibinfo{publisher}{Wiley}, \bibinfo{year}{2000}).

\bibitem[{\citenamefont{Box and Tiao}(1973)}]{BT}
\bibinfo{author}{\bibfnamefont{R.}~\bibnamefont{Box}} \bibnamefont{and}
  \bibinfo{author}{\bibfnamefont{G.~C.} \bibnamefont{Tiao}},
  \emph{\bibinfo{title}{Bayesian Inference in Statistical Analysis}}
  (\bibinfo{publisher}{Addison-Wesley, Sydney}, \bibinfo{year}{1973}).

\end{thebibliography}

\end{document}